\documentclass{aastex62}
\usepackage{amssymb,amsmath}
\usepackage{color}

\submitjournal{ApJ}

\shorttitle{Boris-HLLD}
\shortauthors{Matsumoto et al.}

\begin{document}

\title{A new HLLD Riemann solver with Boris correction for reducing Alfv\'en speed}

\correspondingauthor{Tomoaki Matsumoto}
\email{matsu@hosei.ac.jp}

\author[0000-0002-8125-4509]{Tomoaki Matsumoto}
\affiliation{Faculty of Sustainability Studies, Hosei University, Fujimi, Chiyoda-ku, Tokyo 102-8160, Japan}
\affiliation{Department of Astrophysical Sciences, Princeton University, 4 Ivy Lane, Princeton, NJ 08544, USA}
\affiliation{RIKEN Center for Computational Science (R-CCS), 7-1-26 Minatojima-minami-machi, Chuo-ku, Kobe, Hyogo 650-0047, Japan}

\author{Takahiro Miyoshi}
\affiliation{Department of Physical Science, Graduate School of Science, Hiroshima University, 1-3-1 Kagamiyama, Higashihiroshima, Hiroshima 739-8526, Japan}

\author[0000-0003-3882-3945]{Shinsuke Takasao}
\affiliation{Department of Physics, Nagoya University, Nagoya, Aichi, 464-8602, Japan}

\begin{abstract}
A new Riemann solver is presented for the ideal magnetohydrodynamics (MHD) equations with the so-called Boris correction. The Boris correction is applied to reduce wave speeds, avoiding an extremely small timestep in MHD simulations. The proposed Riemann solver, Boris-HLLD, is based on the HLLD solver. As done by the original HLLD solver, (1) the Boris-HLLD solver has four intermediate states in the Riemann fan when left and right states are given, (2) it resolves the contact discontinuity, Alfv\'en waves, and fast waves, and (3) it satisfies all the jump conditions across shock waves and discontinuities except for slow shock waves.  The results of a shock tube problem indicate that the scheme with the Boris-HLLD solver captures contact discontinuities sharply and it exhibits shock waves without any overshoot when using the minmod limiter. The stability tests show that the scheme is stable when $|u| \lesssim 0.5c$ for a low Alfv\'en speed ($V_A \lesssim c$), where $u$, $c$, and $V_A$ denote the gas velocity, speed of light, and Alfv\'en speed, respectively. For a high Alfv\'en speed ($V_A \gtrsim c$), where the plasma beta is relatively low in many cases, the stable region is large, $|u| \lesssim (0.6-1) c$. We discuss the effect of the Boris correction on physical quantities using several test problems. The Boris-HLLD scheme can be useful for problems with supersonic flows in which regions with a very low plasma beta appear in the computational domain.
\end{abstract}


\section{Introduction} \label{sec:intro}

Magnetohydrodynamics (MHD) simulations are widely used in astrophysics and space sciences. When MHD equations are solved using explicit methods, a high Alfven speed often arises in a region where the plasma beta is low, and it requires a small timestep, which makes long-term calculations very difficult. The Alfv\'en speed becomes high when the magnetic field is strong or the gas density is low. An extremely low density requires an extremely small timestep, which greatly increases the number of timesteps and thus the computational burden.  Such difficulties often arise when gravity is taken into account in MHD simulations \citep[e.g.,][]{Matsumoto04}. In MHD simulations of the magnetospheres of strongly magnetized planets, the fast Alfv\'en wave due to a planetary dipole field causes the same difficulties \citep[e.g.,][]{Toth12}.

In order to avoid a very small timestep due to an extremely large Alfv\'en speed, two approaches are commonly adopted. The first approach is to artificially reduce the Lorentz force \citep{Rempel09}. This method directly weakens the magnetic effects. The second approach is to impose a variable inertia so that the time rate of change of the flow velocity is reduced as the magnetic field strengthens, leading to slow Alfv\'en and magnetosonic waves. The latter approach is called the Boris correction \citep{Boris70}. With this correction, the speeds of the Alfv\'en and magnetosonic waves are bounded by the speed of light, $c$, which can be set to an artificially low value.  Moreover, using the simplified version of the Boris correction,the steady-state solutions are independent of $c$. Both approaches are included in the semi-relativistic MHD equations \citep{Gombosi02}.

Methods that utilize a reduced Alfv\'en speed have been used not only in steady-state simulations but also in time-dependent simulations of star formation \citep[e.g.,][]{Allen03}, accretion disks \citep[e.g.,][]{Miller00,Parkin14}, and solar physics \citep[e.g.,][]{Rempel17}, and in space sciences \citep[e.g.,][]{Lyon04,Toth12}. Shock-capturing schemes usually employ the total variation diminishing (TVD) approach, but conventional Riemann solvers, such as the Lax-Friedrichs scheme and the Harten-Lax-van Leer (HLL) scheme \citep{Harten83}, have also been adopted. A high-resolution, semi-relativistic Riemann solver that resolves many shocks and discontinuities is desirable. The HLLD Riemann solver \citep{Miyoshi05} is one of the most widely used high-resolution schemes, being used in simulation codes such as \texttt{SFUMATO} \citep{Matsumoto07,Matsumoto17}, \texttt{Athena++} \citep{Stone19,Takasao18}, and \texttt{PLUTO} \citep{Mignone12}. It resolves the contact discontinuity, Alfv\'en waves, and fast waves. 
 Although \citet{Parkin14} used the HLLD solver with the Boris correction, a simplified approach, in which the numerical flux is not compatible with the Boris correction, was adopted.

In this paper, we propose a scheme that incorporates the Boris correction into the HLLD Riemann solver. The rest of this paper is organized as follows. In Section~\ref{sec:HLLDBoris}, the HLLD solver with the Boris correction is derived. The results of numerical tests are presented in Section~\ref{sec:test}. Finally, a summary of the main results and a discussion are presented in Section~\ref{sec:summary}.

\section{Incorporation of Boris correction into HLLD solver}
\label{sec:HLLDBoris}
\subsection{Governing equations}
\label{sec:governingeq}
The governing equations of the semi-relativistic equation with the Boris simplification are given as follows \citep{Gombosi02},
\begin{equation}
\frac{\partial \mathbf{U}}{\partial t}
+ \nabla \cdot \mathbb{F}
= 0,
\label{eq:mhdpde}
\end{equation}
\begin{equation}
\mathbf{U} = 
\left(
\begin{array}{c}
\rho\\
(1+V_A^2/c^2)\rho \mathbf{u}\\
\mathbf{B}\\
e
\end{array}
\right),
\label{eq:U}
\end{equation}
\begin{equation}
\mathbb{F} = 
\left(
\begin{array}{c}
\rho \mathbf{u}\\
\rho \mathbf{u}\mathbf{u} + p_{T} - \mathbf{B}\mathbf{B} \\
\mathbf{u}\mathbf{B}-\mathbf{B}\mathbf{u}\\
(e + p_{T} ) \mathbf{u}  -
(\mathbf{B}\cdot\mathbf{u})\mathbf{B}
\end{array}
\right) ,
\label{eq:F}
\end{equation}
\begin{gather}
\mathbf{u} =(u,v,w)^T, \\
\mathbf{B} =(B_x, B_y, B_z)^T, \\
V_A^2 = \frac{\left|\mathbf{B}\right|^2}{\rho},\\
e = \rho \frac{\left|\mathbf{u}\right|^2 }{ 2} + \frac{p}{\gamma - 1} +
\frac{\left|\mathbf{B}\right|^2}{2}, \\
p_{T} = p + \frac{\left|\mathbf{B}\right|^2}{2},
\end{gather}
where $\mathbf{U}$, $\mathbb{F}$, $\rho$, $\mathbf{u}$, $p$, $p_T$, $e$, $\mathbf{B}$, $V_A$, and $c$ are the state vector, flux, density, velocity, pressure, total pressure, total energy, magnetic field, Alfv\'en speed, and speed of light, respectively.  The superscript $T$ denotes the transpose of a vector. Hereafter, we refer to this formulation as the Boris correction.
The difference between this formulation and the original MHD equations is the factor $\left(1+V_A^2/c^2\right)$ in the three components of momentum (see Equation~(\ref{eq:U})).  This factor indicates that inertia becomes large where the Alfv\'en speed is high compared to the reduced speed of light $c$.
For later convenience, we define $\rho_A$ as,
\begin{equation}
\rho_A = \frac{|\mathbf{B}|^2}{c^2},
\end{equation}
and write the three components of momentum as
$\left(\rho + \rho_A\right)\mathbf{u}$, indicating that $\rho_A$ is the extra inertia caused by the magnetic fields.

For one-dimensional problems, the governing equations reduce to,
\begin{equation}
\frac{\partial \mathbf{U}}{\partial t}
+\frac{\partial \mathbf{F}}{\partial x}
= 0
\label{eq:mhdpde1d}
\end{equation}
\begin{equation}
\mathbf{U} = 
\left(
\begin{array}{c}
\rho\\
(\rho + \rho_A)u\\
(\rho + \rho_A)v\\
(\rho + \rho_A)w\\
B_y\\
B_z\\
e
\end{array}
\right) ,
\label{eq:U1d}
\end{equation}
\begin{equation}
\mathbf{F} = 
\left(
\begin{array}{c}
\rho u\\
\rho u^2 + p_{T} - B_x^2 \\
\rho uv - B_x B_y \\
\rho uw - B_x B_z \\
B_y u - B_x v\\
B_z u - B_x w\\
(e + p_{T} ) u  -
(\mathbf{B}\cdot\mathbf{u})B_x
\end{array}
\right).
\label{eq:F1d}
\end{equation}

\begin{figure}
  \begin{center}
    \epsscale{0.5}
    \plotone{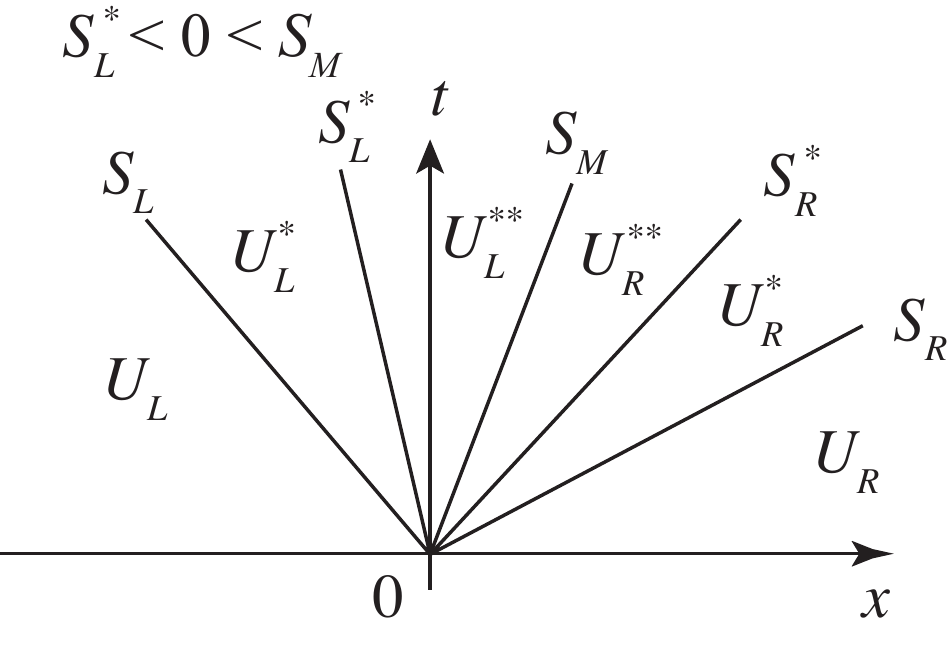}
  \end{center}
  \caption{Schematic diagram of the Riemann fan for the Boris-HLLD Riemann solver. This figure shows the case of $S_L^* < 0 < S_M$.  The Riemann fan consists of the four intermediate states $\mathbf{U}_L^*$, $\mathbf{U}_L^{**}$, $\mathbf{U}_R^{**}$, and $\mathbf{U}_R^*$ for given states $\mathbf{U}_L$ and $\mathbf{U}_R$. The states are separated by the five wave speeds $S_L$, $S_L^*$, $S_M$, $S_R^*$, and $S_R$.
  \label{riemannfan.pdf}}
\end{figure}

\subsection{Assumptions for Riemann solver}
\label{sec:assumption_Boris_HLLD}
As is the case for the original HLLD solver, the HLLD solver with the Boris correction (hereafter Boris-HLLD solver) is assumed to have four intermediate states, $\mathbf{U}_L^*$, $\mathbf{U}_L^{**}$, $\mathbf{U}_R^{**}$, and $\mathbf{U}_R^*$, for the given left and right states, $\mathbf{U}_L$ and $\mathbf{U}_R$ (Figure~\ref{riemannfan.pdf}).  The states are separated by the five wave speeds $S_L$, $S_L^*$, $S_M$, $S_R^*$, and $S_R$.  The wave speeds $S_L$ and $S_R$ are the highest wave speeds, usually the speeds of the fast waves, $S_L^*$ and $S_R^*$ are the speeds of the Alfv\'en waves, and $S_M$ is the speed of an entropy wave.  
Hereafter, we denote the physical variables in the intermediate states with the superscripts $*/**$ and subscripts $L/R$; for example, the density in the state
$\mathbf{U}^*_L$ is $\rho^*_L$.
Similarly, the fluxes, $\mathbf{F}_L$, $\mathbf{F}_L^*$, $\mathbf{F}_L^{**}$, $\mathbf{F}_R^{**}$, $\mathbf{F}_R^{*}$, and $\mathbf{F}_R$, are defined. For example, the flux in the state $\mathbf{U}^*_L$ is
  $\mathbf{F}_L^* = \mathbf{F}(\rho^*_L, \mathbf{u}^*_L, \mathbf{B}^*_L, p_{TL}^*, e^*_L)$, where the function $\mathbf{F}(\rho, \mathbf{u}, \mathbf{B}, p_T, e)$ is given by Equation~(\ref{eq:F1d}).
Throughout all the states, $B_x$ is constant because of the divergence-free feature of the magnetic field.

We construct the following jump condition for each wave,
\begin{gather}
S_\alpha \mathbf{U}^*_\alpha - \mathbf{F}^*_\alpha = S_\alpha \mathbf{U}_\alpha - \mathbf{F}_\alpha,
\label{eq:jumps}\\
S^*_\alpha \mathbf{U}^{**}_\alpha - \mathbf{F}^{**}_\alpha = S^*_\alpha \mathbf{U}^*_\alpha - \mathbf{F}^*_\alpha,
\label{eq:jumpsa}\\
S_M \mathbf{U}^{**}_R - \mathbf{F}^{**}_R = S_M \mathbf{U}^{**}_L - \mathbf{F}^{**}_L,
\label{eq:jumpsm}
\end{gather}
where $\alpha = L$ or $R$.  
Equations~(\ref{eq:jumps}), (\ref{eq:jumpsa}), and (\ref{eq:jumpsm}) are the jump conditions across $S_\alpha$, $S^*_\alpha$, and $S_M$, respectively.  The intermediate states are derived based on these jump conditions. 

First, we assume that $u$ and $p_T$ are constant in the Riemann fan, yielding, 
\begin{equation}
u^*_L = u^{**}_L = u^*_R = u^{**}_R = S_M,
\label{eq:uconst}
\end{equation}
and 
\begin{equation}
p_{TL}^* = p_{TL}^{**} = p_{TR}^* = p_{TR}^{**}.
\label{eq:pconst}
\end{equation}
This assumption is the same as that in the original HLLD solver. Equation~(\ref{eq:uconst}) and the mass conservation component of the jump condition across $S^*_\alpha$ (Equation~(\ref{eq:jumpsa})) give, 
\begin{equation}
\rho^{**}_\alpha = \rho^*_\alpha,
\label{eq:rhoam}
\end{equation}
for $S^*_\alpha \ne S_M$.  This relation is the same as that for the original HLLD solver.

The $x$ momentum component of the three jump conditions across $S_M$ (Equation~(\ref{eq:jumpsm})) and $S^*_\alpha$ (Equation~(\ref{eq:jumpsa})) gives,
\begin{equation}
\left|\mathbf{B}^*_L\right|^2 = \left|\mathbf{B}^*_R\right|^2
= \left|\mathbf{B}^{**}_L\right|^2 = \left|\mathbf{B}^{**}_R\right|^2 = \rho_A c^2, 
\label{eq:Bconstant}
\end{equation}
indicating that $\rho_A$ is constant throughout the intermediate states. 

\subsection{Evaluation of $\mathbf{U}^*_\alpha$}
Considering the jump conditions across $S_\alpha$ (Equation~(\ref{eq:jumps})), the mass conservation component yields,
\begin{equation}
\rho^*_\alpha = \rho_\alpha \frac{S_\alpha - u_\alpha}{S_\alpha - S_M}.
\end{equation}
The $x$ momentum and mass conservation components give,
\begin{equation}
S_M = \frac{(S^\prime_R - u_R)\rho_Ru_R - (S^\prime_L - u_L)\rho_Lu_L - p_{TR} + p_{TL}}
{(S^\prime_R - u_R)\rho_R - (S^\prime_L - u_L)\rho_L},
\end{equation}
\begin{equation}
p^*_{T\alpha}
=
\frac{(S^\prime_R-u_R)\rho_Rp_{TL} - (S^\prime_L-u_L)\rho_Lp_{TR} +
  \rho_L\rho_R(S^\prime_R-u_R)(S^\prime_L-u_L)(u_R-u_L)}
{(S^\prime_R-u_R)\rho_R-(S^\prime_L-u_L)\rho_L},
\end{equation}
where
\begin{equation}
S^\prime_\alpha 
= \left(1+\frac{V_A^2}{c^2}\right) S_\alpha
= \left(1+\frac{\rho_A}{\rho_\alpha}\right) S_\alpha
= \left(1+\frac{ |\mathbf{B}^*|^2 }{\rho_\alpha c^2}\right) S_\alpha.
\label{eq:sdlr}
\end{equation}
The $y$ components of the momentum and induction equations of the jump conditions give,
\begin{equation}
v^*_\alpha = v_\alpha - 
B_{y\alpha} B_x 
\frac{S_M-u_\alpha}
{\rho_\alpha(S^\prime_\alpha-u_\alpha)(S_\alpha-S_M)-B_x^2} ,
\label{eq:valr}
\end{equation}
\begin{equation}
B_{y\alpha}^* =
B_{y\alpha}
\frac{ \rho_\alpha(S^\prime_\alpha-u_\alpha)(S_\alpha-u_\alpha)-B_x^2}
{\rho_\alpha(S^\prime_\alpha-u_\alpha)(S_\alpha-S_M)-B_x^2}.
\label{eq:byalr}
\end{equation}
Similarly, the $z$ components give,
\begin{equation}
w^*_\alpha = w_\alpha - 
B_{z\alpha} B_x 
\frac{S_M-u_\alpha}
{\rho_\alpha(S^\prime_\alpha-u_\alpha)(S_\alpha-S_M)-B_x^2} ,
\end{equation}
\begin{equation}
B_{z\alpha}^* =
B_{z\alpha}
\frac{ \rho_\alpha(S^\prime_\alpha-u_\alpha)(S_\alpha-u_\alpha)-B_x^2}
{\rho_\alpha(S^\prime_\alpha-u_\alpha)(S_\alpha-S_M)-B_x^2} .
\label{eq:bzalr}
\end{equation}
From the energy conservation component,
\begin{equation}
e^*_\alpha = \frac
{(S_\alpha-u_\alpha)e_\alpha-p_{T\alpha} u_\alpha+p_{T\alpha}^* S_M+B_x
(\mathbf{u}_\alpha\cdot\mathbf{B}_\alpha - \mathbf{u}^*_\alpha\cdot\mathbf{B}^*_\alpha)
}
{S_\alpha-S_M} 
\end{equation}
is derived. 

\subsection{Evaluation of $\mathbf{U}^{**}_\alpha$}

Considering the jump conditions across $S^*_\alpha$ (Equation~(\ref{eq:jumpsa})), the $y$ and $z$ components of the momentum and induction equations are satisfied for arbitrary $v^*_\alpha$, $w^*_\alpha$, $v^{**}_\alpha$, $w^{**}_\alpha$, $B_{y\alpha}^*$, $B_{z\alpha}^*$, $B_{y\alpha}^{**}$, and $B_{z\alpha}^{**}$ when the following equations hold,
\begin{equation}
S^*_\alpha = 
\frac{1}{2}
\left[
\left( 1+ \gamma_A^2 \right)S_M \mp 
\sqrt{ \left(1-\gamma_A^2 \right)^2 S_M^2 + 4 \gamma_A^2 V_{A,x}^2 }
\right],
\label{eq;salr}
\end{equation}
\begin{equation}
V_{A,x}^2 = \frac{B_x^2}{\rho^*_\alpha},
\end{equation}
\begin{equation}
\gamma_A = \frac{1}{\sqrt{1+V_A^2/c^2}}
= \frac{1}{\sqrt{1+ |\mathbf{B}^*|^2/(\rho^*_\alpha c^2)}}
= \frac{1}{\sqrt{1+ \rho_A/\rho^*_\alpha}},
\end{equation}
where the sign $\mp$ corresponds to $\alpha=L$ and $R$. The subscript $\alpha$ is omitted in $V_{A,x}$ and $\gamma_A$ for simplicity, although they depend on the right and left states. For $V_A \ll c$, $S^*_\alpha \rightarrow S_M \mp V_{A,x}$ holds, which coincides with $S^*_\alpha$ in the original HLLD solver. For the special case of $S_M = 0$ and $V_A \gg c$, $S^*_\alpha \rightarrow \mp c$, indicating that the wave speed is bounded by $c$. For the more general case of $V_A \gg c$, $S^*_\alpha \rightarrow S_M$ or 0. Equation~(\ref{eq;salr}) was also derived by \citet{Gombosi02}. 
From the energy component of the jump conditions,
\begin{equation}
e^{**}_\alpha = e^*_\alpha + \frac{B_x (\mathbf{u}^*_\alpha\cdot\mathbf{B}^*_\alpha-\mathbf{u}^{**}_\alpha\cdot\mathbf{B}^{**}_\alpha)}{S^*_\alpha-S_M}
\end{equation}
is obtained.  The remaining components (mass conservation and $x$ component of the momentum) hold with those variables.

Considering the jump condition across $S_M$ (Equation~(\ref{eq:jumpsm})), the $y$ and $z$ components of the momentum and induction equations yield,
\begin{gather}
v^{**}_L = v^{**}_R = v^{**},\\
w^{**}_L = w^{**}_R = w^{**},\\
B^{**}_{yL} = B^{**}_{yR} = B^{**}_y,\\
B^{**}_{zL} = B^{**}_{zR} = B^{**}_z,
\end{gather}
where the same relationships hold in the original HLLD solver.  These variables are also satisfied with the energy component of the jump condition.

The jump condition through the Riemann fan,
\begin{equation}
(S_R - S^*_R) \mathbf{U}^*_R + 
(S^*_R - S_M) \mathbf{U}^{**}_R + 
(S_M - S^*_L) \mathbf{U}^{**}_L + 
(S^*_L - S_L) \mathbf{U}^*_L
-S_R\mathbf{U}_R+S_L\mathbf{U}_L
+\mathbf{F}_R-\mathbf{F}_L = 0,
\label{eq:jcond_full}
\end{equation}
and the jump condition across $S^*_\alpha$ (Equation~(\ref{eq:jumpsa})) are rewritten as, 
\begin{equation}
 - S^*_R \mathbf{U}^*_R + 
(S^*_R - S_M) \mathbf{U}^{**}_R + 
(S_M - S^*_L) \mathbf{U}^{**}_L + 
S^*_L \mathbf{U}^*_L
+\mathbf{F}^*_R-\mathbf{F}^*_L = 0.
\label{eq:jcond_reduce}
\end{equation}
The $y$ and $z$ components of the momentum and induction equations of Equation~(\ref{eq:jcond_reduce}) yield,
\begin{equation}
v^{**} = \frac
{S^*_Rv^*_R(\rho^*_R+\rho_A) - S^*_Lv^*_L(\rho^*_L+\rho_A) 
-(\rho^*_Rv^*_R - \rho^*_Lv^*_L)S_M + B_x( B_{yR}^* - B_{yL}^*)
}
{(S^*_R-S_M)(\rho^*_R+\rho_A) + (S_M-S^*_L)(\rho^*_L+\rho_A) },
\label{eq:va}
\end{equation}
\begin{equation}
w^{**} = \frac
{S^*_Rw^*_R(\rho^*_R+\rho_A) - S^*_Lw^*_L(\rho^*_L+\rho_A) 
-(\rho^*_Rw^*_R - \rho^*_Lw^*_L)S_M + B_x( B_{zR}^* - B_{zL}^*)
}
{(S^*_R-S_M)(\rho^*_R+\rho_A) + (S_M-S^*_L)(\rho^*_L+\rho_A) },
\end{equation}
\begin{equation}
B_y^{**} = \frac
{(S^*_R-S_M)B_{yR}^* + (S_M-S^*_L)B_{yL}^* + B_x(v^*_R-v^*_L)  }
{S^*_R-S^*_L},
\end{equation}
\begin{equation}
B_z^{**} = \frac
{(S^*_R-S_M)B_{zR}^* + (S_M-S^*_L)B_{zL}^* + B_x(w^*_R-w^*_L)  }
{S^*_R-S^*_L}.
\label{eq:bza}
\end{equation}
For $B_x = 0$, the Alfv\'en wave propagating in the $x$ direction does not exist, and $\mathbf{U}^*_L$ and $\mathbf{U}^*_R$ are adopted as the intermediate states instead of $\mathbf{U}^{**}_L$ and $\mathbf{U}^{**}_R$, respectively, as in the original HLLD solver.

\subsection{Approximation of $\rho_A$}

Once $\rho_A$ is obtained, all the components of the intermediate states can be obtained.  Formally, $\rho_A$ should be derived from Equations~(\ref{eq:Bconstant}), (\ref{eq:sdlr}), (\ref{eq:byalr}), and (\ref{eq:bzalr}).  However, the simultaneous equations are too complicated to obtain a simple formula for numerical computation.  Here, we assume that $\rho_A$ is approximated by the so-called HLL average,
\begin{equation}
\overline{\mathbf{U}} = \frac{S_R \mathbf{U}_R - S_L \mathbf{U}_L - \mathbf{F}_R + \mathbf{F}_L}{S_R - S_L}.
\end{equation}
The $B_y$ and $B_z$ components are,
\begin{equation}
\left(
\begin{array}{c}
\overline{B}_y\\
\overline{B}_z
\end{array}
\right) =
\frac{1}{S_R-S_L}
\left[
S_R
\left(
\begin{array}{c}
B_{yR}\\
B_{zR}
\end{array}
\right)
-
S_L
\left(
\begin{array}{c}
B_{yL}\\
B_{zL}
\end{array}
\right)
-
\left(
\begin{array}{c}
B_{yR} u_R - B_x v_R\\
B_{zR} u_R - B_x w_R
\end{array}
\right)
+
\left(
\begin{array}{c}
B_{yL} u_L - B_x v_L\\
B_{zL} u_L - B_x w_L
\end{array}
\right)
\right].
\end{equation}
By using $\overline{B}_y$ and $\overline{B}_z$, we adopt 
\begin{equation}
\rho_A = \frac{B_x^2 + \overline{B}_y^2 + \overline{B}_z^2}{c^2}.
\end{equation}

\subsection{$S_L$ and $S_R$}
\label{sec:slsr}
The wave speeds $S_L$ and $S_R$ are the highest wave speeds in the two directions, and specify the expansion of the Riemann fan. The following speed settings are usable in practice,
\begin{align}
S_L &= \min\left(u_L-c_{\mathrm{fast},L}^B, u_R-c_{\mathrm{fast},R}^B\right),
\label{eq:sl}\\
S_R &= \max\left(u_L+c_{\mathrm{fast},L}^B, u_R+c_{\mathrm{fast},R}^B\right),
\label{eq:sr}
\end{align}
where $c_{\mathrm{fast},L}^B$ and $c_{\mathrm{fast},R}^B$ are the speeds of the fast wave in the left and right states, respectively, for the Boris correction.  For the governing equations (Equations~(\ref{eq:mhdpde1d})--(\ref{eq:F1d})), \citet{Gombosi02} solved the speed of the fast wave for the case of $\mathbf{u}=0$,
\begin{gather}
c_\mathrm{fast}^B = \frac{\gamma_A}{\sqrt{2}} 
\sqrt{a^2 + V_A^2 + \sqrt{\left(a^2+V_A^2\right)^2 - 4 a^2 V_{A,x}^2}},
\label{eq:cfastslow}\\
a^2 =\frac{\gamma p}{\rho}.
\end{gather}
For $V_A \ll c$, the wave speed coincides with that of the classical fast wave.  For $V_A \rightarrow \infty$, $c_\mathrm{fast}^B \rightarrow c$, and it is bounded by $c$. We adopt Equation~(\ref{eq:cfastslow}) for $c_{\mathrm{fast},L}^B$ and $c_{\mathrm{fast},R}^B$ in the numerical computations, substituting the primitive variables of the corresponding right and left states.

The appropriate order of the wave speeds, $S_L \leq S^*_L$ and $S^*_R \leq S_R$, is not guaranteed when Equations~(\ref{eq:sl}), (\ref{eq:sr}), and (\ref{eq:cfastslow}) are adopted as $S_L$ and $S_R$. We therefore arrange the order of the wave speeds using the following correction,
\begin{align}
  S_L &= \min\left(S_L, S^*_L\right), &
  S_R &= \max\left(S_R, S^*_R\right).
\label{eq:slrmodifiy}
\end{align}

In addition, the denominators of Equations~(\ref{eq:valr})--(\ref{eq:bzalr}) can become zero. For very small denominators, e.g., ${\rho_\alpha(S^\prime_\alpha-u_\alpha)(S_\alpha-S_M)-B_x^2}< \epsilon p^*_{T\alpha}$, the wave speed $S_L$ is reduced and $S_R$ is increased by a small value in order to avoid a zero denominator, where $\epsilon\, (=10^{-6})$ denotes a small factor. This corresponds to an extension of the Riemann fan. The small value for reducing/increasing $S_{L/R}$ is set to $(S_R-S_L) \epsilon$, where $S_R - S_L$ is the total width of the Riemann fan. 

When $S_L$ and $S_R$ are corrected according to the rearrangement of the order of wave speeds and the correction for zero denominators, the dependent variables $\rho^*_\alpha$, $\rho_A$, $S_M$, $S^*_\alpha$, and $S^\prime_\alpha$ should be recalculated in principle. However, we only recalculate $S^\prime_\alpha$ because the recalculation of all the dependent variables does not guarantee an appropriate ordering of wave speeds and non-zero denominators. Other treatments may be possible depending on the implementation.

The correction factor $\gamma_A$ in Equation~(\ref{eq:cfastslow}) produces an over-correction of the fast wave speed; it is reduced to $c_\mathrm{fast}^B$ even for the case of $c_\mathrm{fast} \lesssim c$, where $c_\mathrm{fast}$ $(=c_\mathrm{fast}^B\gamma_A^{-1})$ denotes the speed of the classical fast wave. The speed of the classical fast wave is therefore adopted if $c_\mathrm{fast} \le c$; otherwise, Equation~(\ref{eq:cfastslow}) is adopted. This switching of the fast wave is effective for the case where both the gas velocity and fast wave speed are low. The effects of switching are discussed in Section~\ref{sec:stability_linearwaves}.

\subsection{Numerical flux}

Given the left and right states $\mathbf{U}_L$ and $\mathbf{U}_R$, the intermediate states $\mathbf{U}_L^*$, $\mathbf{U}_L^{**}$, $\mathbf{U}_R^{**}$, and $\mathbf{U}_R^*$ are obtained using the primitive variables derived above. When the fluxes in the right and left states $\mathbf{F}_L$ and $\mathbf{F}_R$ are given, the numerical fluxes are sequentially obtained with the jump conditions for all the intermediate states,
\begin{align}
\mathbf{F}_L^* &= \mathbf{F}_L + S_L(\mathbf{U}_L^* - \mathbf{U}_L),&
\mathbf{F}_L^{**} &= \mathbf{F}_L^{*} + S_L^*(\mathbf{U}_L^{**} - \mathbf{U}_L^*),\nonumber\\
\mathbf{F}_R^* &= \mathbf{F}_R + S_R(\mathbf{U}_R^* - \mathbf{U}_R),&
\mathbf{F}_R^{**} &= \mathbf{F}_R^{*} + S_R^*(\mathbf{U}_R^{**} - \mathbf{U}_R^*).
\label{eq:fluxjumpcond}
\end{align}
Alternatively, the following method may be easier than Equation~(\ref{eq:fluxjumpcond}),
\begin{align}
\mathbf{F}_L^* &= \mathbf{F}(\rho^*_L, \mathbf{u}^*_L, \mathbf{B}^*_L, p_{TL}^*, e^*_L),&
\mathbf{F}_L^{**} &= \mathbf{F}(\rho^{**}_L, \mathbf{u}^{**}_L, \mathbf{B}^{**}_L, p_{TL}^{**}, e^{**}_L),\nonumber\\
\mathbf{F}_R^{**} &= \mathbf{F}(\rho^{**}_R, \mathbf{u}^{**}_R, \mathbf{B}^{**}_R, p_{TR}^{**}, e^{**}_R), &
\mathbf{F}_R^* &= \mathbf{F}(\rho^*_R, \mathbf{u}^*_R, \mathbf{B}^*_R, p_{TR}^*, e^*_R),
\end{align}
where the function $\mathbf{F}(\rho, \mathbf{u}, \mathbf{B}, p_T, e)$ is given by Equation~(\ref{eq:F1d}).
 As in the original HLLD solver, the numerical flux is switched according to the wave speeds, 
\begin{equation}
  \mathbf{F}^\mathrm{Boris-HLLD} =
  \begin{cases}
    \mathbf{F}_L   & \text{if } S_L > 0,\\
    \mathbf{F}_L^* & \text{if } S_L \le 0 < S_L^*,\\
    \mathbf{F}_L^{**} & \text{if } S_L^* \le 0 < S_M,\\
    \mathbf{F}_R^{**} & \text{if } S_M \le 0 < S_R^*,\\
    \mathbf{F}_R^{*} & \text{if } S_R^* \le 0 < S_R,\\
    \mathbf{F}_R & \text{if }  S_R \le 0.
  \end{cases}
\label{eq:f_hlld}
\end{equation}
For example, for the case shown in Figure~\ref{riemannfan.pdf}, $\mathbf{F}_L^{**}$ is adopted as the numerical flux.

The intermediate states and waves derived in Sections~\ref{sec:assumption_Boris_HLLD} to \ref{sec:slsr} coincide with those in the original HLLD solver in the limit of $V_A \ll c$. The solutions are therefore expected to coincide with those obtained by the original HLLD solver in this limit.

\section{Numerical tests}
\label{sec:test}

\subsection{Implementation}

For the test calculations, we use \texttt{SFUMATO} \citep{Matsumoto07}, in which the Boris-HLLD solver is implemented.  Adaptive mesh refinement is switched off (i.e., uniform grids are utilized).  The scheme has second-order accuracy in time and space with the predictor-corrector method and the Monotonic Upwind Scheme for Conservation Laws (MUSCL), respectively. The minmod limiter is adopted as a slope limiter in the MUSCL unless explicitly mentioned. Hyperbolic divergence cleaning \citep{Dedner02} is adopted for the $\nabla\cdot\mathbf{B}$ treatment. The specific heat is set to $\gamma=5/3$ for all problems.

Incorporating the Boris-HLLD scheme is easy; the numerical flux of the original HLLD solver is replaced by that of the Boris-HLLD solver, given in Equation~(\ref{eq:f_hlld}). 
The state vector $\mathbf{U}$ is modified according to Equation~(\ref{eq:U}).
The timestep is determined based on the Courant-Friedrichs-Lewy (CFL) condition, and the reduced fast wave speed is adopted for the three-dimensional case,
\begin{equation}
\Delta t = C_\mathrm{CFL} \min_{i,j,k} \left[\left(\frac{| u |+ c_\mathrm{fast}^{B,\mathrm{max}} }{\Delta x}+\frac{| v | + c_\mathrm{fast}^{B,\mathrm{max}} }{\Delta y}+\frac{| w | + c_\mathrm{fast}^{B,\mathrm{max}}}{\Delta z}\right)^{-1}_{i,j,k}\right],
\label{eq:CFLcondition3d}
\end{equation}
where $C_\mathrm{CFL}$ denotes the CFL number (typically 0.7), the subscripts $(i,j,k)$ specify a cell, and $(\Delta x, \Delta y, \Delta z)$ denote the cell widths. The wave speed $c_\mathrm{fast}^{B,\mathrm{max}}$ is the maximum speed of the reduced fast wave. The reduced fast wave speed is given by Equation~(\ref{eq:cfastslow}); its maximum value is given by,
\begin{equation}
c_\mathrm{fast}^{B,\mathrm{max}} = \gamma_A \sqrt{a^2 + V_A^2 }.
\label{eq:cfastmax}
\end{equation}
This CFL condition is the same as that for the classical MHD solver except for the factor $\gamma_A = \left(1+V_A^2/c^2 \right)^{-1/2}$ in Equation~(\ref{eq:cfastmax}).
For the one-dimensional case, Equation~(\ref{eq:CFLcondition3d}) reduces to,
\begin{equation}
\Delta t = C_\mathrm{CFL} \min_i \left(\frac{\Delta x_i}{| u_{i} |+ c_{\mathrm{fast},i}^{B,\mathrm{max}} }\right).
\label{eq:CFLcondition1d}
\end{equation}

As shown later, the proposed scheme becomes unstable in a certain situation. 
In order to confirm that the instability comes from the formulation of the Boris simplification, not from the discretization of the Boris-HLLD solver introduced in Section~\ref{sec:HLLDBoris}, we incorporate the Boris correction also into the HLL Riemann solver. The numerical flux of the HLL solver with the Boris correction (hereafter Boris-HLL solver) is given by
\begin{equation}
\mathbf{F}^\mathrm{Boris-HLL}=\frac{S_R \mathbf{F}_L - S_L \mathbf{F}_R + S_R S_L (\mathbf{U}_R - \mathbf{U}_L) }{S_R - S_L},
\end{equation}
where
\begin{gather}
S_L = \min\left(u_L-c_{\mathrm{fast},L}^B, u_R-c_{\mathrm{fast},R}^B, \lambda_{A,L}^{-}, \lambda_{A,R}^{-}, 0\right),\label{eq:HLLSL}\\
S_R = \max\left(u_L+c_{\mathrm{fast},L}^B, u_R+c_{\mathrm{fast},R}^B, \lambda_{A,L}^{+}, \lambda_{A,R}^{+}, 0\right),\label{eq:HLLSR}\\
\lambda_{A,\alpha}^{\pm} = \frac{1}{2} \left[\left(1+\gamma_{A,\alpha}^2\right)u_\alpha \pm \sqrt{\left(1-\gamma_{A,\alpha}^2\right)^2 u_\alpha^2+4 \gamma_{A,\alpha}^2 V_{A,x,\alpha}^2} \right],\label{eq:HLLlambda}\\
V_{A,x,\alpha}^2 = \frac{B_x^2}{\rho_\alpha},\\
\gamma_{A,\alpha}^2 = (1+V_{A,\alpha}^2/c^2)^{-1},
\end{gather}
for $\alpha = L$ or $R$. The signal speed of the Alfv\'en wave is denoted by $\lambda_{A,\alpha}^\pm$. Equations~(\ref{eq:HLLSL}) and (\ref{eq:HLLSR}) are the same as Equations~(\ref{eq:sl}) and (\ref{eq:sr}), but they also arrange the order of the wave speeds in a way similar to Equation~(\ref{eq:slrmodifiy}). This arrangement is necessary for the Alfv\'en and sound waves to be stable in  stability tests in Section~\ref{sec:stability_linearwaves}. The switching of the fast wave described in Section~\ref{sec:slsr} is implemented because it is effective also in the Boris-HLL solver.
The Boris-HLL solver is implemented in \texttt{SFUMATO}. The scheme is the same as the Boris-HLLD scheme except for the numerical flux,  providing a fair comparison between the Boris-HLLD and Boris-HLL solvers.

In order to confirm that the instability does not arise from the implementation of the code, we use \texttt{Athena++} \citep{Stone19} for comparison. The \texttt{Athena++} code also adopts a scheme with second-order accuracy in time and space with the predictor-corrector method and MUSCL, respectively. The van Leer limiter is adopted as a slope limiter in the MUSCL. For the $\nabla\cdot\mathbf{B}$ treatment, the constraint transport method is adopted, conserving the initial $\nabla\cdot\mathbf{B}$ within machine accuracy \citep{Stone09}. 

\subsection{Shock tube problem}
\label{sec:shocktube}
The standard MHD shock tube problem proposed by \citet{BrioWu88} is solved. In the computational domain $x \in [-5, 5]$ with 256 mesh points, the initial state is set to $\rho=1$, $B_y=1$, and $p=1$ for $x < 0$, and $\rho=0.125$, $B_y=-1$, and $p=0.1$ for $x \ge 0$. Throughout the computational domain, $B_x=0.75$, $B_z=0$, and $\mathbf{u}=0$ are constant. The Alfv\'en speed is $V_A = 1.25$ for $x<0$ and 3.5 for $x>0$. The wave speeds are reduced considerably for $x>0$ when the speed of light is set to $c=3$ or 2. For comparison, the original HLLD solver and the Boris-HLL solver are used in addition to the proposed Boris-HLLD scheme. 

\begin{figure}
\epsscale{1}
\plottwo{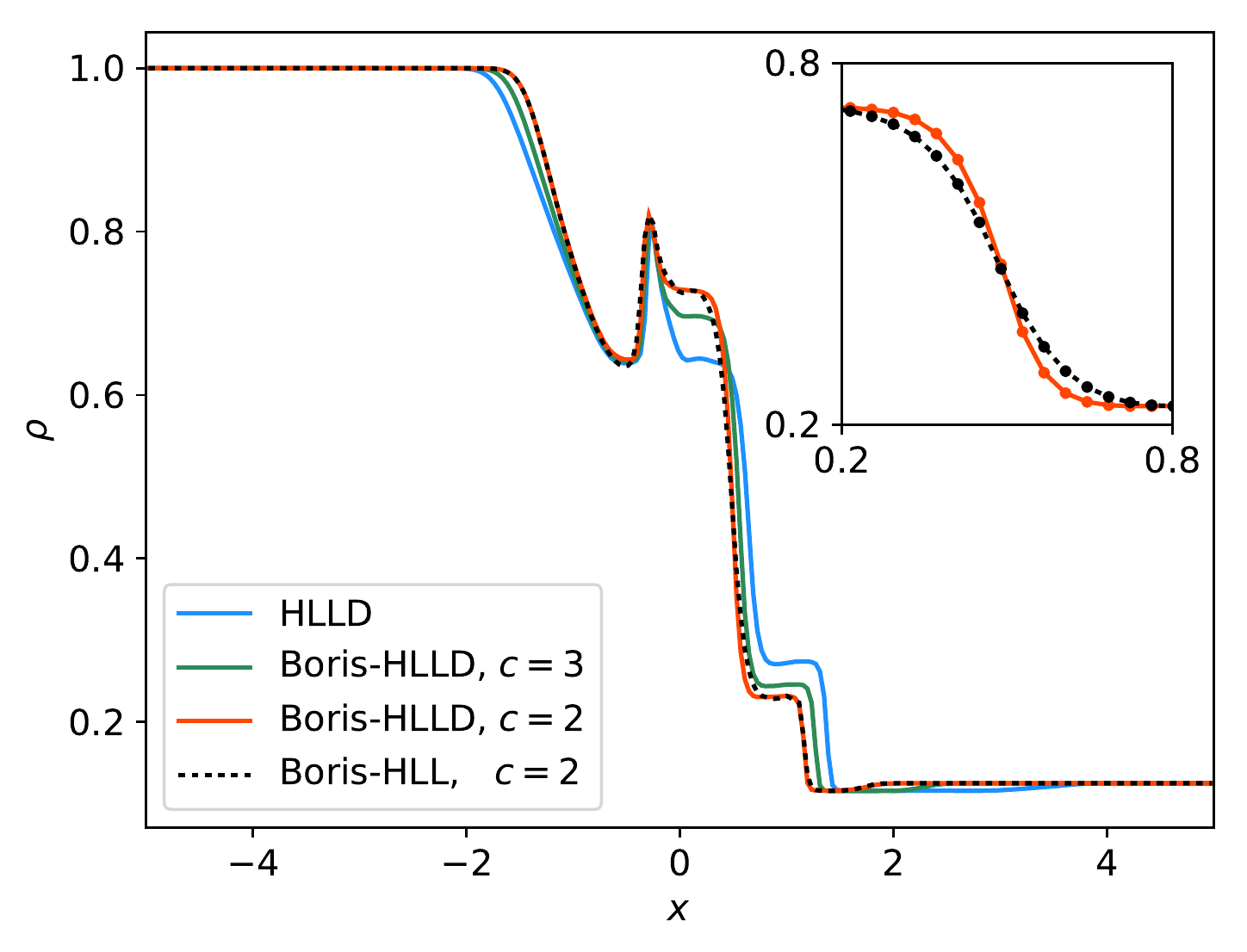}{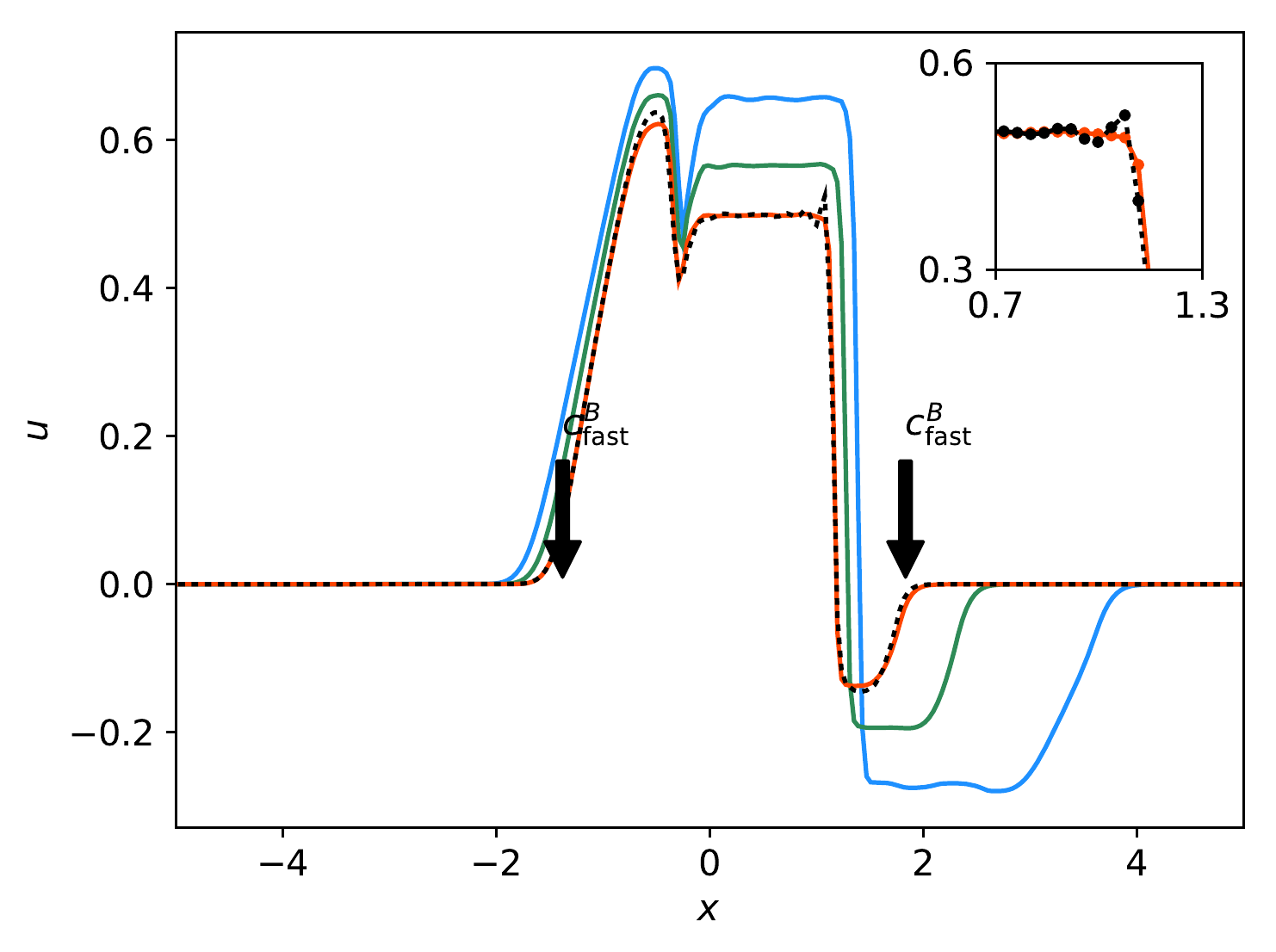}
\plottwo{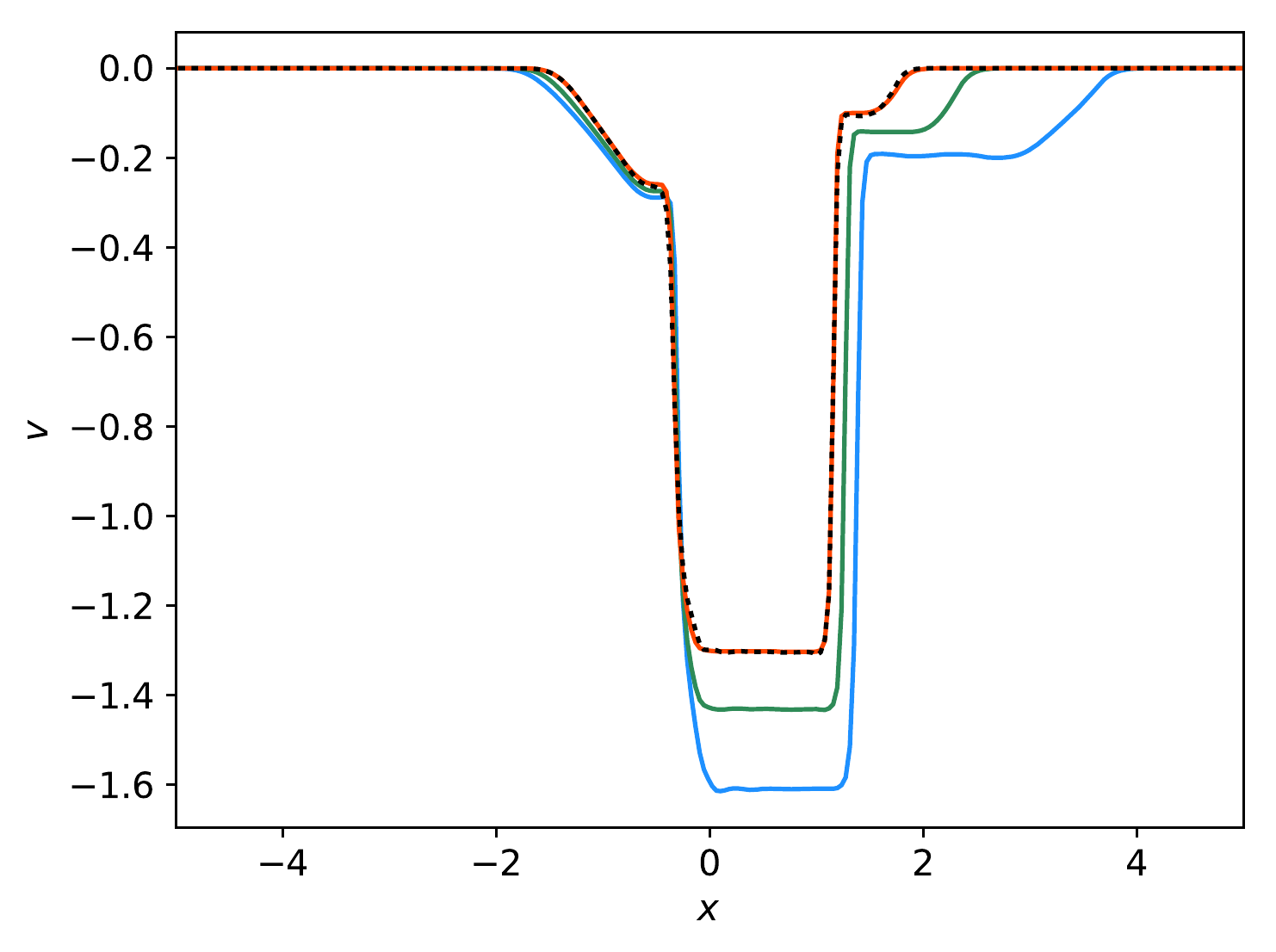}{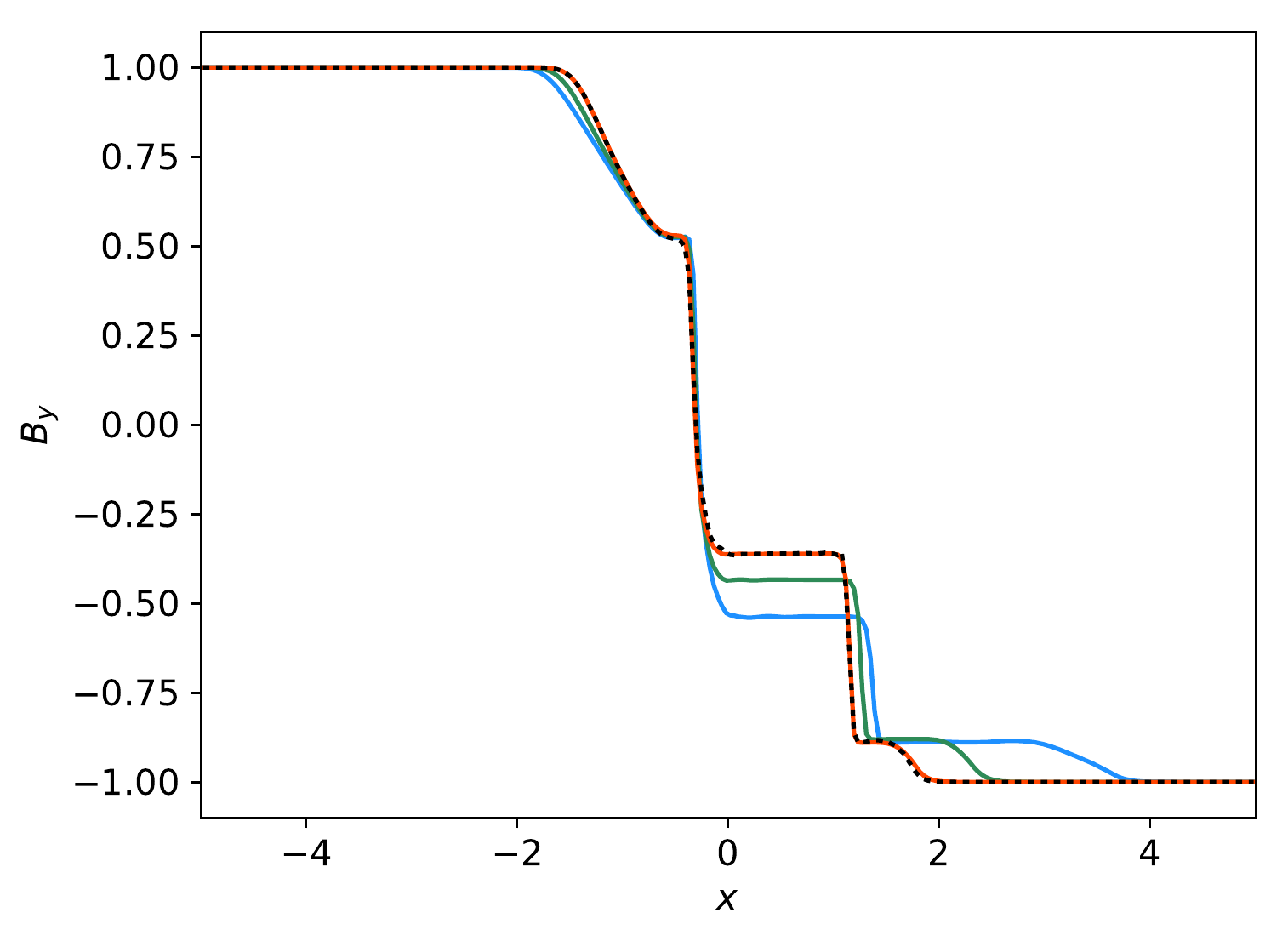}
\plottwo{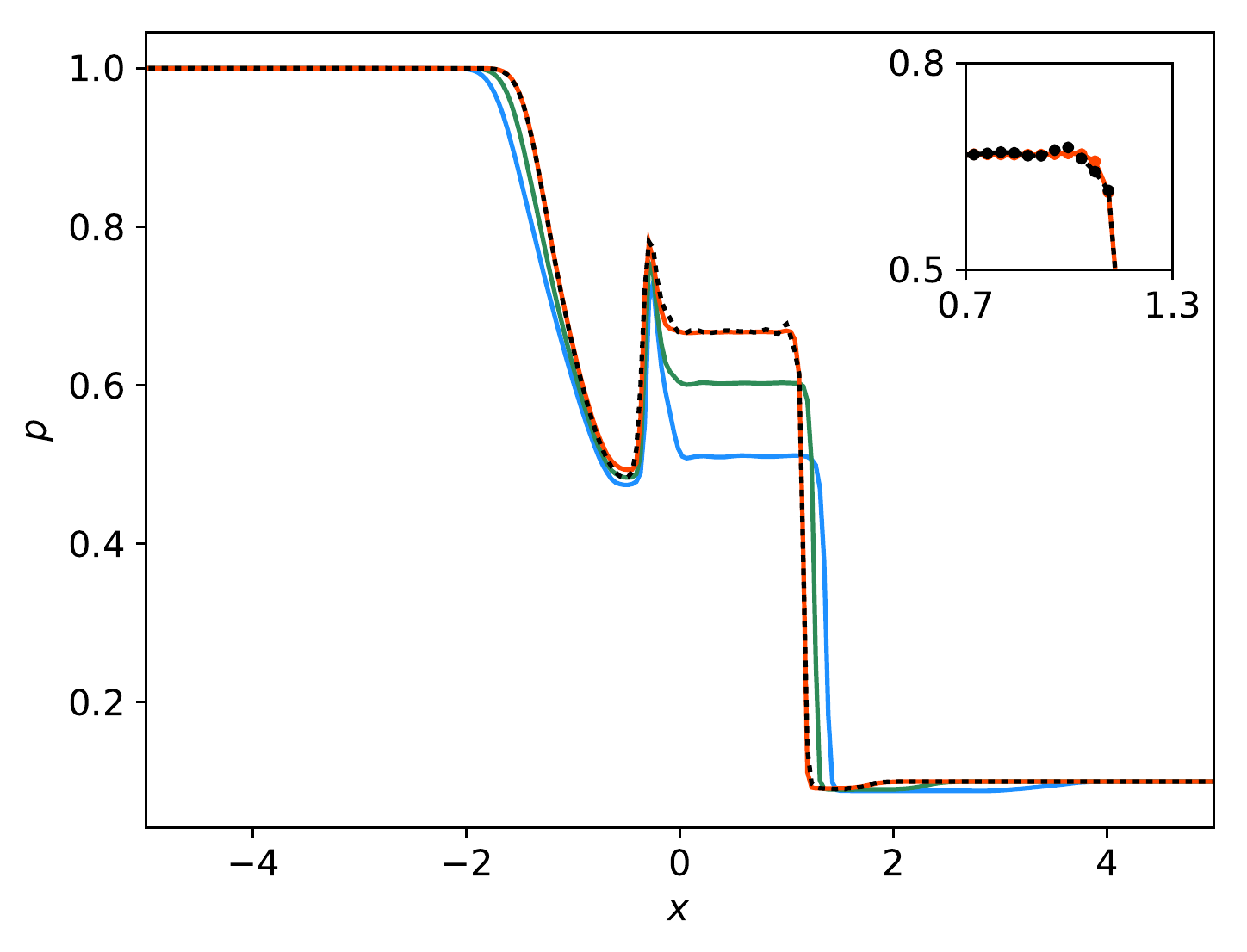}{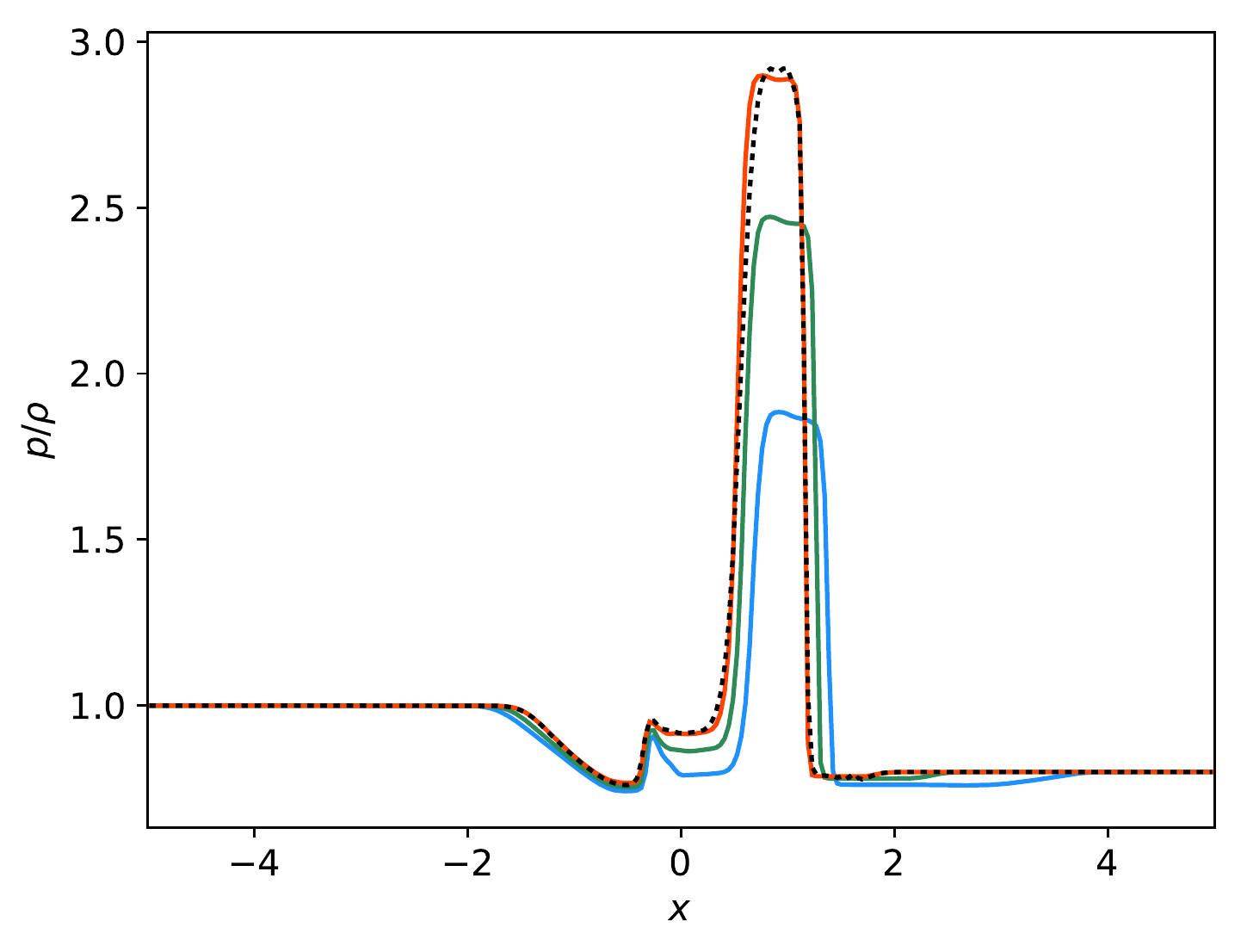}
\caption{Shock tube problem solved by the HLLD scheme, the Boris-HLLD scheme with different speeds of light ($c=3, 2$), and the Boris-HLL scheme with $c=2$. The distributions of $\rho$, $u$, $v$, $B_y$, $p$, and $p/\rho$ are shown at $t=1$. The arrows in the top right panel indicate the expected location of the fast wave obtained from Equation~(\ref{eq:cfastslow}) for $c=2$. The insets show a comparison of the solutions between the Boris-HLLD and Boris-HLL schemes with $c=2$ for the contact discontinuity (top left panel) and slow shock (top right and bottom left panels). The black circles in the insets denote the mesh points.
\label{shocktube.pdf}}
\end{figure}

Figure~\ref{shocktube.pdf} shows the distributions of physical variables in the shock tube problem. As shown, the shock waves and discontinuities are sharply resolved without any overshoot by the Boris-HLLD solver. Because of the Boris correction, the wave speeds are reduced as $c$ decreases. The fast rarefaction wave slows down, and the location of the wave front is in good agreement with that expected from Equation~(\ref{eq:cfastslow}), as indicated by the arrows in the figure. Both velocity components ($u$ and $v$) decrease because of the extra inertia $\rho_A$, as can be seen for the fast rarefaction wave and the slow shock at $x\simeq 1.5$. The density, pressure, and magnetic field also change. The decrease in the velocity jump leads to a decrease in the density jump at the slow shock. In contrast, the pressure jump increases, resulting in a considerable increase in the temperature of the post-shock gas, as indicated by the $p/\rho$ distribution. This test problem demonstrates the influence of the Boris correction on the solution.

The solution is not considerably affected by the Boris correction for $x < 0$. This is because the Alfv\'en speed is considerably lower than the speed of light in this region. The solution converges to that of the original HLLD solver as $c$ increases.

Setting $c$ to a low value decreases the number of timesteps. For the shock tube problem examined here, the original HLLD scheme requires 142 timesteps, while the Boris-HLLD scheme requires 105 and 94 timesteps for $c=3$ and 2, respectively. 

Figure~\ref{shocktube.pdf} also compares the solutions between the Boris-HLLD and Boris-HLL solvers in the case of $c=2$. The solutions obtained with the two solvers are in good agreement. However, the Boris-HLLD solver exhibits sharper profiles of the contact discontinuity in the distributions of $\rho$ and $p/\rho$ than the Boris-HLL solver does because the Boris-HLLD solver resolves a contact discontinuity.  The difference in the sharpness between the two solvers is approximately the same as the difference between the original HLLD and HLL solvers. At the slow shock, the Boris-HLL solver produces small overshoots in the distributions of $u$ and $p$. The size of the overshoots depends on a slope limiter adopted in the MUSCL. When we use the van Leer limiter, which is a steeper limiter than the minmod limiter, both the Boris-HLLD and Boris-HLL scheme show overshoots at the slow shock, but the Boris-HLLD scheme exhibits a considerably smaller overshoot than the Boris-HLL scheme. The solution with the Boris-HLLD solver is more accurate than that with the Boris-HLL solver because it exhibits a sharp contact discontinuity and a slow shock with a small or no overshoot.

\subsection{Linear Alfv\'en waves}
\label{sec:linear_alfvenwave}
We consider linear Alfv\'en waves propagating parallel to a uniform magnetic field $B_0$ with a uniform density $\rho_0$. The linear analysis of the one-dimensional governing equations (Equations~(\ref{eq:mhdpde1d})--(\ref{eq:F1d})) leads to the following eigen mode of the Alfv\'en wave,
\begin{equation}
\left(
\begin{array}{c}
v\\
B_y
\end{array}
\right) =
\left(
\begin{array}{c}
\gamma_A V_A\\
-B_0
\end{array}
\right)\delta_\mathrm{pert} \sin ( k x - \omega t),
\label{eq:linear_alfvenwave}
\end{equation}
where
\begin{gather}
\gamma_A = \left(1+V_A^2/c^2\right)^{-1/2},\\
V_A = \frac{B_0}{\sqrt{\rho_0}},\\
\frac{\omega}{k} = \gamma_A V_A.
\end{gather}

The initial condition was constructed according to Equation~(\ref{eq:linear_alfvenwave}) with $t=0$, $\rho_0=1$, $B_0 = 1$, and $\delta_\mathrm{pert} = 10^{-5}$. The classical Alfv\'en speed is therefore $V_A =1$. The initial condition also has $\rho=\rho_0$, $p=1$, $u=w=0$, $B_x = B_0$, and $B_z =0$. The wavelength is set to $L=1$ ($k=2\pi$). The computational domain is $x \in [-L/2, L/2]$ with a uniform grid with 128 mesh points. The periodic boundary condition is imposed at $x=-L/2$ and $L/2$. The calculation is terminated at $t_\mathrm{last}=1$, which is the wave crossing time for $c=\infty$. 

\begin{figure}
\epsscale{1}
\plottwo{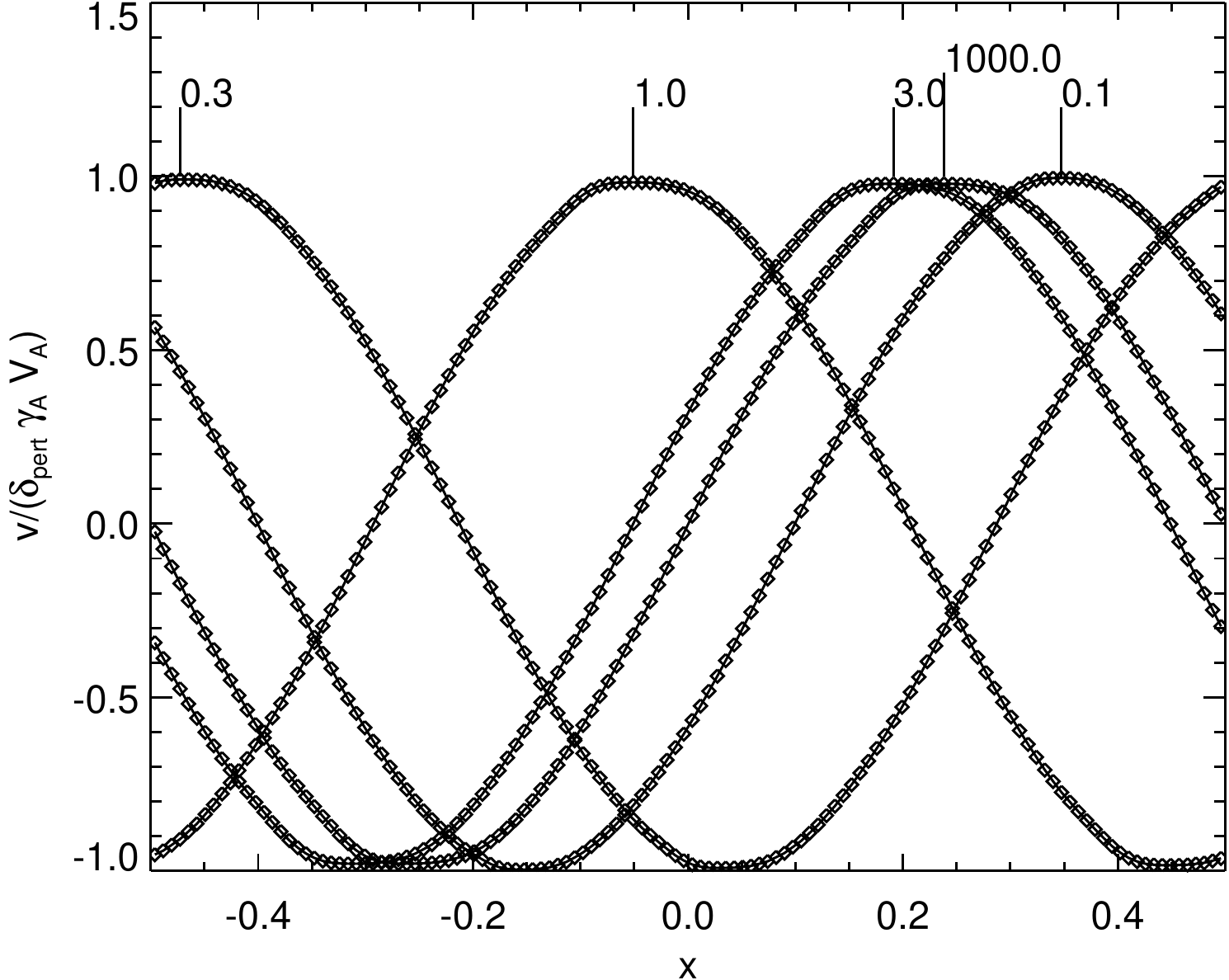}{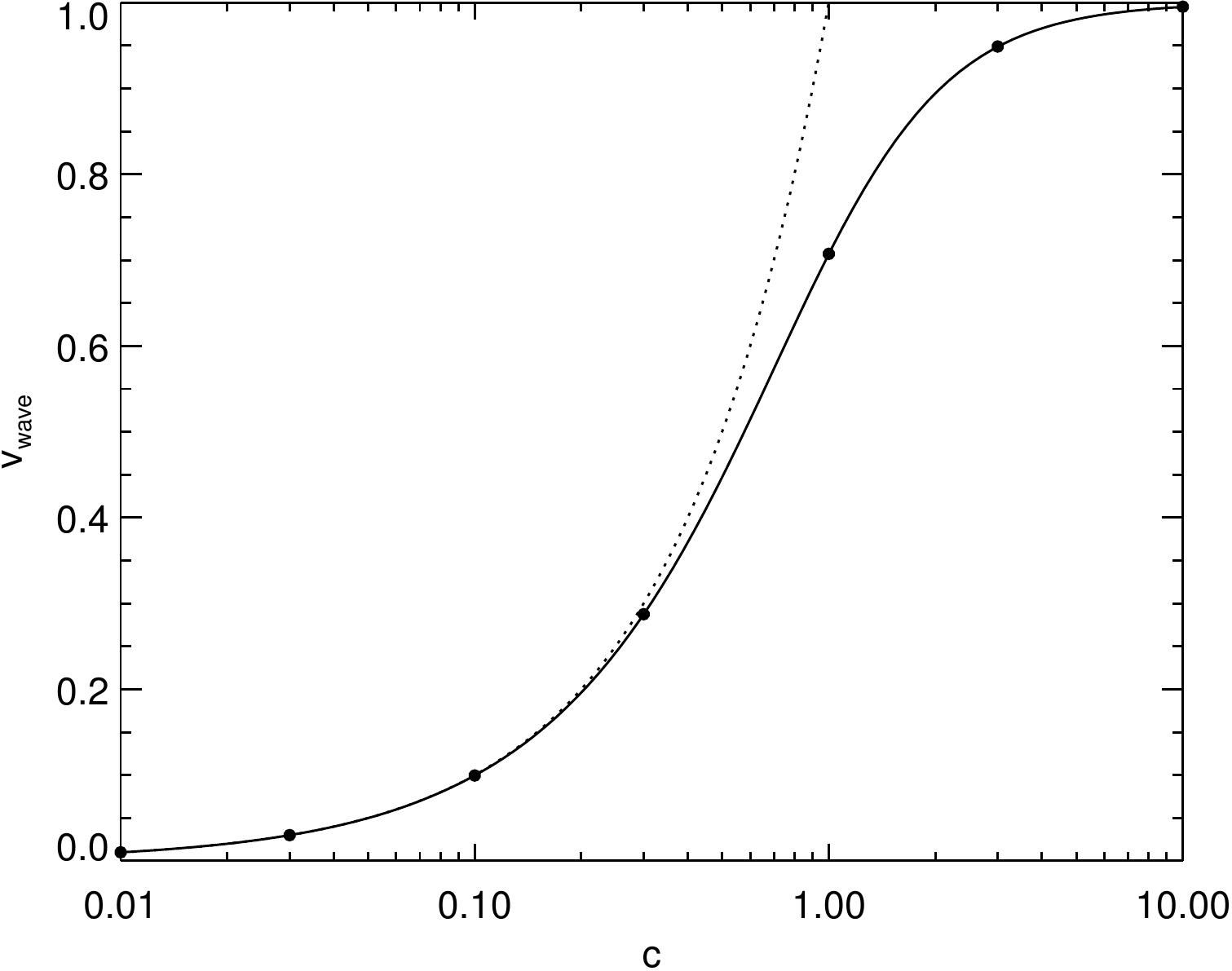}
\caption{
Propagation of a linear Alfv\'en wave with different speeds of light $c$. Left panel: profiles of $v$ as a function of $x$ for $c=0.1, 0.3, 1.0, 3.0,$ and 1000 at $t = t_\mathrm{last}$ $(=1)$. The ordinate is normalized by the initial amplitude $\delta_\mathrm{pert}\gamma_A V_A$. The speed of light $c$ is labeled for each line.  The diamonds show the mesh points. Right panel: propagation velocity $v_\mathrm{wave}$ as a function of $c$. The dots denote the numerical values. The solid and dotted lines are the relationships of $v_\mathrm{wave} = \gamma_A V_A$ and $v_\mathrm{wave} =c$, respectively. The horizontal axis, $c$, is plotted on a logarithmic scale.
\label{alfvenwave.pdf}
}
\end{figure}

Figure~\ref{alfvenwave.pdf} (left panel) shows the profiles of $v$ for different $c$ values at $t=1$. The wave with $c=1000$ propagates a distance of one wavelength. The travel distance becomes shorter for a lower $c$. Figure~\ref{alfvenwave.pdf} (right panel) shows the wave velocity measured in the calculations. In order to measure the velocity for each wave, the travel distance of the wave was evaluated with a phase offset of the first mode of the Fourier transform on the wave profile at $t=t_\mathrm{last}$. The measured wave velocities are in agreement with the theoretical values, which are shown by the solid line. The wave velocities are limited by $c$. Those with high $c$ asymptotically approach the classical Alfv\'en speed ($V_A =1$ in this case).

\subsection{Stability of linear waves}
\label{sec:stability_linearwaves}
The stability of the Alfv\'en wave, sound wave, and fast magnetosonic wave is investigated numerically. The propagation of a linear wave is calculated, and the increase in the amplitude of the wave is measured for each parameter. We use the amplification factor of the wave as an indicator of instability.

Alfv\'en waves propagating along a magnetic field are considered, as in Section~\ref{sec:linear_alfvenwave}, but the bulk motion of the gas is also taken into account.  An Alfv\'en wave with a small amplitude in the moving gas has the following eigen mode for the governing equations (Equations~(\ref{eq:mhdpde1d})--(\ref{eq:F1d})),
\begin{equation}
\left(
\begin{array}{c}
v\\
B_y
\end{array}
\right) =
\left(
\begin{array}{c}
\lambda_A - u_0\\
-B_0
\end{array}
\right)\delta_\mathrm{pert} \sin ( k x - \omega t),
\label{eq:stab_alfvenwave}
\end{equation}
where $\lambda_A$ is the signal speed of the Alfv\'en wave, given by,
\begin{equation}
\lambda_A = \frac{1}{2} \left[\left(1+\gamma_A^2\right)u_0 + \sqrt{\left(1-\gamma_A^2\right)^2 u_0^2+4 \gamma_A^2 V_A^2} \right].
\label{eq:lambda_a}
\end{equation}
The initial condition was constructed according to Equation~(\ref{eq:stab_alfvenwave}). We set $\rho=\rho_0=1$, $p=p_0$, $w=0$, $B_x = B_0$, $B_z =0$, and $\delta_\mathrm{pert} = 10^{-4}$ at $t=0$. The wavelength is set to $L=1$ ($k=2\pi$). The computational domain is $x \in [-L/2, L/2]$ with a uniform grid with 128 mesh points.  The periodic boundary condition is imposed at $x=-L/2$ and $L/2$. The calculation is terminated at $t_\mathrm{last}=1$.

\begin{figure}
\epsscale{1}
\plottwo{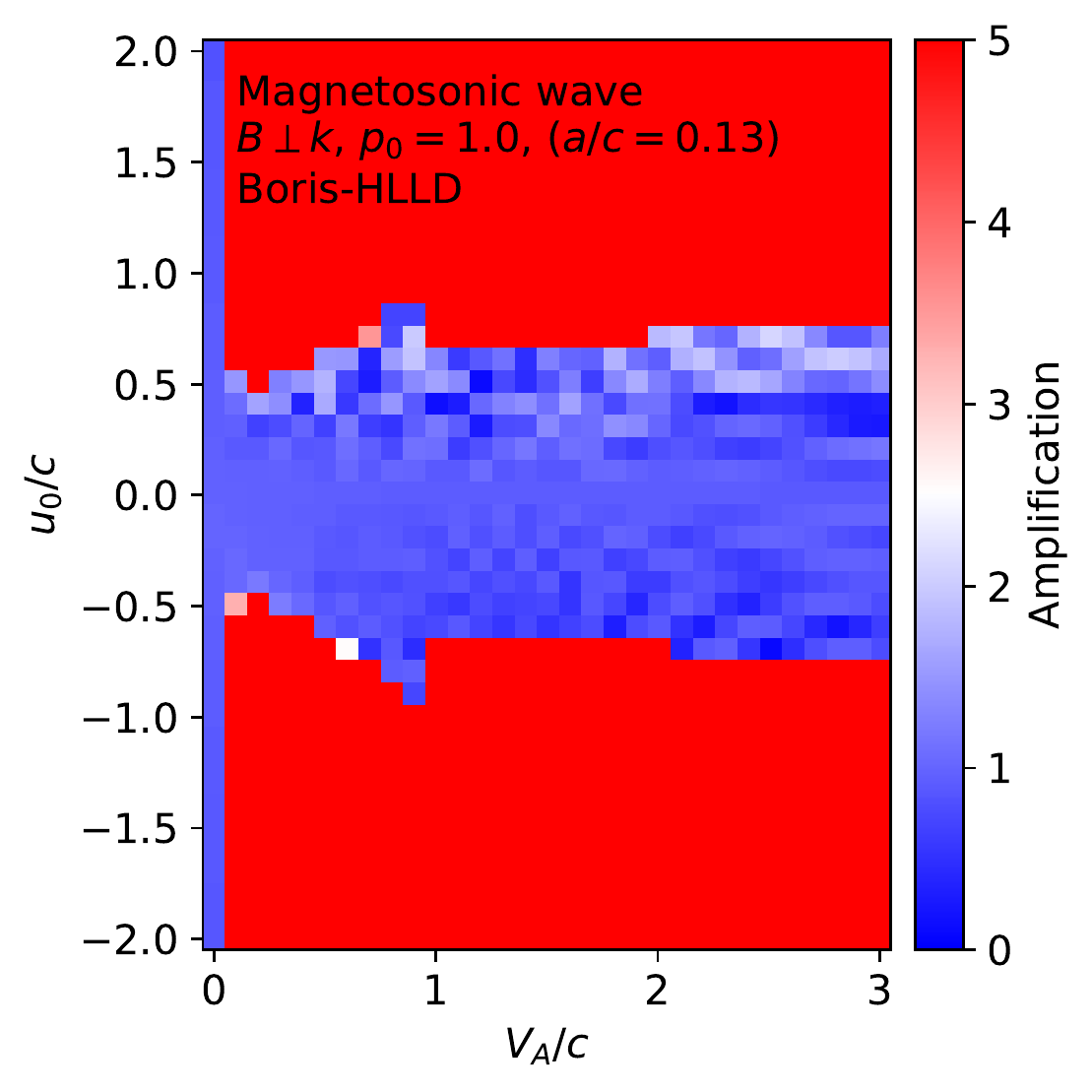}{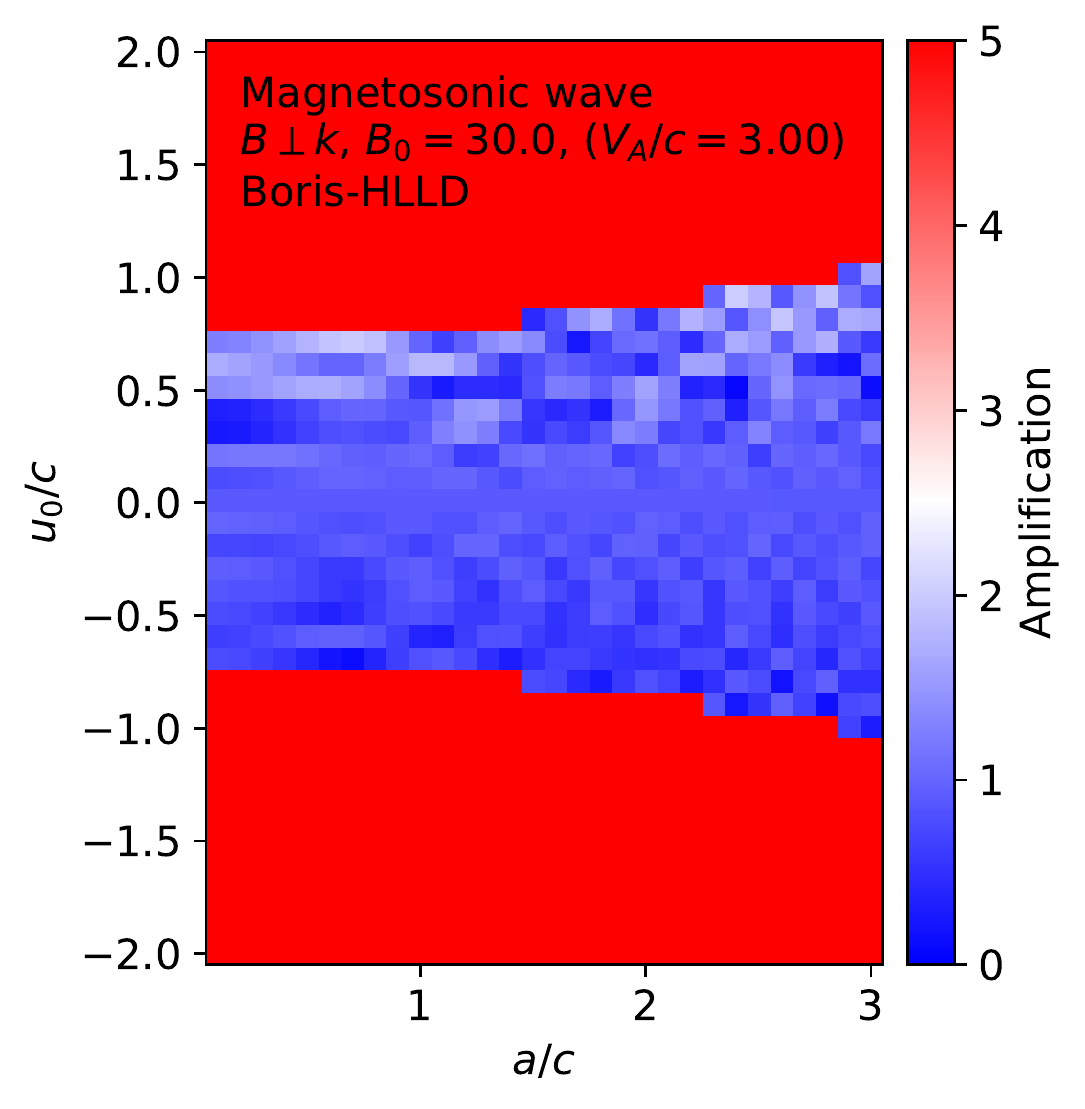}
\plottwo{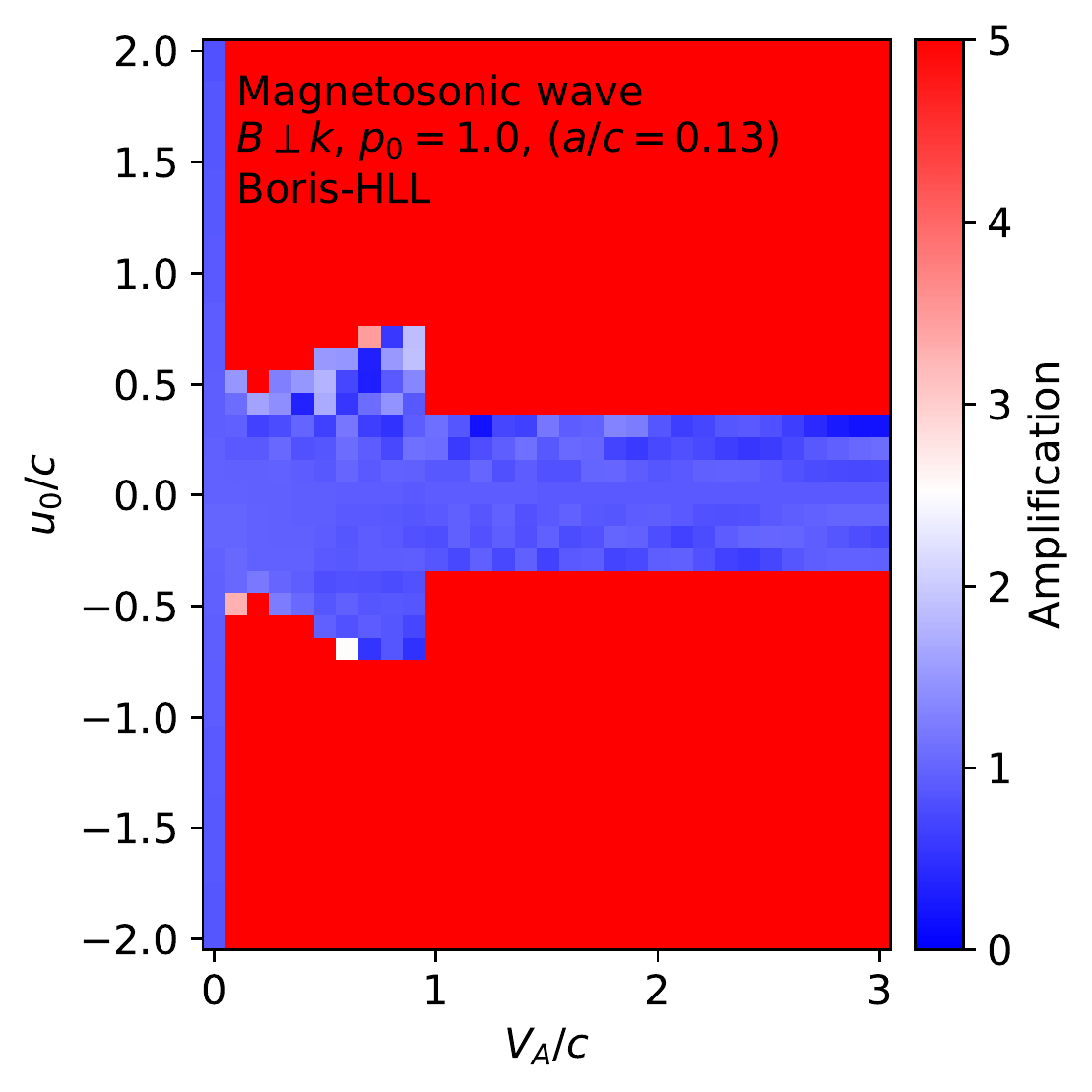}{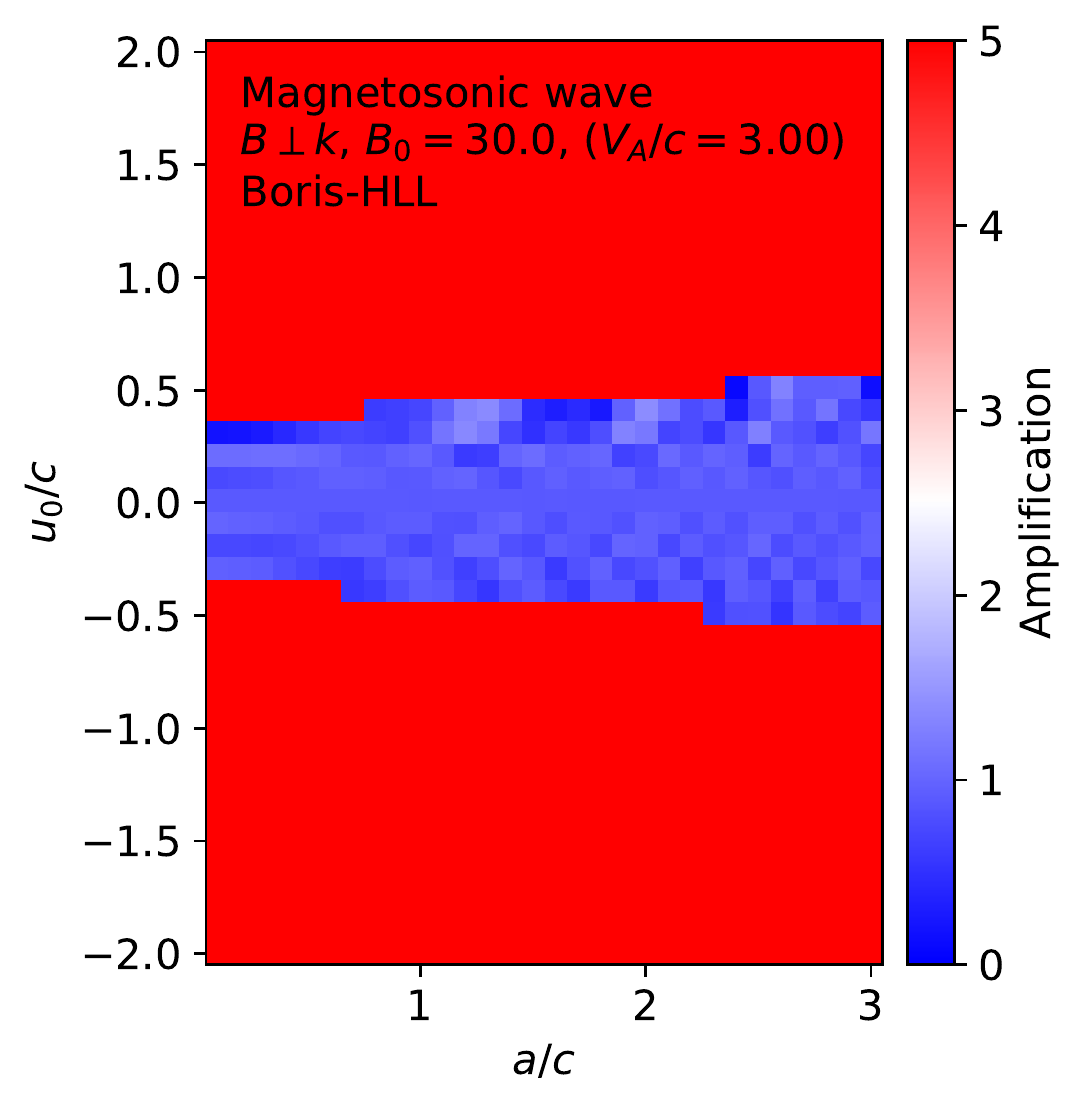}
\caption{
Amplification factor of magnetosonic waves perpendicular to the magnetic field at $t=1$ for the Boris-HLLD scheme (top panels) and the Boris-HLL scheme (bottom panels). The amplification factor is measured using the relative amplitude of $\rho$ at $t=1$ with respect to the initial value. The blue and red regions show where the waves are stable and unstable, respectively. In the red region, the amplification factor exceeds 5 or the calculation crashed before $t=1$ because of a negative pressure. The speed of light was set to $c=10$ for all calculations. We set $p_0 = 1$ ($a/c = 0.13$) (left panels) and $B_0=30$ ($V_A/c = 3$) (right panels).
\label{map_alfvenwave.pdf}
}
\end{figure}

We change $u_0$, $B_0$, and $p_0$ in the range of $V_A/c \in [0.1, 3.0]$, $u_0/c \in [-2, 2]$, and $a/c \in [0.1, 3.0]$ in two-dimensional spaces of $V_A/c - u_0/c$ and $a/c - u_0/c$ (see Figure~\ref{map_alfvenwave.pdf} for the parameter space). The speed of light is set to $c=10$. The numerical tests show that all the Alfv\'en waves are stable irrespective of $u_0$, $V_A$, and $a$ for both the Boris-HLLD and Boris-HLL schemes. This is because the wave speed of the eigen mode (Equation~(\ref{eq:lambda_a})) is always real.

In the semi-relativistic MHD formulations proposed by \citet{Gombosi02}, the Alfv\'en wave parallel to the magnetic field is unstable when $u_0 > V_A$. The difference between our result and their result arises from the adopted form of the equations of motion. \citet{Gombosi02} adopted the semi-relativistic equations of motion, which consider the off-diagonal terms in momentum, whereas we used the equations based on the Boris correction, neglecting the off-diagonal terms. Although the equations with off-diagonal terms are physically preferred, those with the Boris correction result in a more stable scheme.

Next, a sound wave propagating along a magnetic field is considered. For $a>V_A$, the sound wave corresponds to the fast wave, and the Alfv\'en wave degenerates to the slow wave.  For $a < V_A$, the Alfv\'en wave degenerates to the fast wave, and the sound wave corresponds to the slow wave. The eigen mode of the sound wave for the MHD equations with the Boris correction (Equations~(\ref{eq:mhdpde1d})--(\ref{eq:F1d})) is expressed as,
\begin{equation}
\left(
\begin{array}{c}
\rho\\
u\\
p
\end{array}
\right) =
\left(
\begin{array}{c}
\rho_0\\
u_0\\
p_0
\end{array}
\right) +
\left(
\begin{array}{c}
\rho_0\\
\lambda_s - u_0\\
\frac{(\gamma -1)u_0(u_0-\lambda_s)^2 + (u_0-\lambda_s)a^2}{\gamma u_0  - \lambda_s}
\end{array}
\right)\delta_\mathrm{pert} \sin (k x - \omega t),
\label{eq:stab_soundwave}
\end{equation}
where $\lambda_s$ is the signal speed of the sound wave, given by,
\begin{equation}
\lambda_s = \frac{1}{2}
\left[
\sqrt{(1-\gamma_A^2)^2u_0^2\gamma^2
  +(\gamma-1)4\gamma_A^2(1-\gamma_A^2)u_0^2
  +4a^2\gamma_A^2} 
+\gamma(1-\gamma_A^2)u_0
+2\gamma_A^2u_0\right].
\label{eq:lambda_s}
\end{equation}
The signal speed $\lambda_s$ is real because $\gamma > 1$ and $\gamma_A < 1$, indicating that the sound wave is always stable. Although this wave is a pure sound wave parallel to the magnetic field, the signal speed depends on the magnetic field through the factor $\gamma_A$. We examine the propagation of sound waves. The settings are the same as those for the Alfv\'en wave test except for the initial perturbation. The numerical tests show that the sound wave is stable irrespective of $a$, $V_A$, and $u_0$ for both the Boris-HLLD and Boris-HLL schemes.

Finally, we consider a magnetosonic wave propagating perpendicular to the magnetic field.  This wave corresponds to the fast wave. Note that no Alfv\'en or sound waves exist perpendicular to the magnetic field. Since the eigen mode of the fast wave is complex for governing equations (Equations~(\ref{eq:mhdpde1d})--(\ref{eq:F1d})), we simply extend the fast wave in the classical MHD, and the initial condition is given by, 
\begin{equation}
\left(
\begin{array}{c}
\rho\\
u\\
B_y\\
p
\end{array}
\right) =
\left(
\begin{array}{c}
\rho_0\\
u_0\\
B_0\\
p_0
\end{array}
\right) +
\left(
\begin{array}{c}
\rho_0\\
\gamma_A c_{\mathrm{fast}\perp}\\
B_0\\
\gamma p_0
\end{array}
\right)\delta_\mathrm{pert} \sin (k x),
\label{eq:stab_fastwave}
\end{equation}
where $c_{\mathrm{fast}\perp}$ is the classical speed of the fast wave propagating perpendicular to the magnetic field, defined as,
\begin{equation}
c_{\mathrm{fast}\perp} = \sqrt{V_A^2 + \frac{\gamma p_0}{\rho_0}}.
\end{equation}
According to Equation~(\ref{eq:cfastslow}), $\gamma_A c_{\mathrm{fast}\perp}$ is the asymptotic speed of the fast wave. We set $B_x=B_z=0$ and $v=w=0$ in the initial condition. The other parameters are the same as those in the previous tests of the Alfv\'en and sound waves.

Figure~\ref{map_alfvenwave.pdf} (top panels) shows the distribution of the amplification factor of the magnetosonic wave at $t=1$ for the Boris-HLLD scheme. The scheme is stable (blue regions) for $|u_0| \lesssim (0.6-1) c$ when the magnetic field is relatively strong ($V_A \gtrsim c$), as shown in left and right panels. The boundary between the stable and unstable regions weakly depends on $V_A/c$ and $a/c$. In the relatively weak magnetic field case ($V_A \lesssim c$), the unstable region extends down to $|u_0| \sim 0.5 c$ (left panel). In the stable (blue) regions, the amplitude is distributed around unity because the initial conditions are not pure eigen modes. The amplitude of the wave oscillates as time proceeds even for a stable wave. Note that the left edge of the left panel corresponds to the non-magnetized case ($V_A/c=0$), and the scheme is stable irrespective of $u_0$ there.

Figure~\ref{map_alfvenwave.pdf} (bottom panels) shows the distribution of the amplification factor for the Boris-HLL scheme for comparison. Both the Boris-HLLD and Boris-HLL schemes are unstable for large $|u_0|/c$, indicating that the instability comes from the governing equations of the Boris correction rather than the discretization of the schemes. The Boris-HLLD scheme shows larger stable regions than the Boris-HLL scheme in the diagrams, indicating that the Boris-HLLD scheme is more stable than the Boris-HLL scheme for the magnetosonic wave.

Figure~\ref{map_alfvenwave.pdf} shows the results in the case of the minmod limiter in the MUSCL. We confirm that the stable regions do not change even when we use the other limiter, e.g., the van Leer limiter, and when we adopt a scheme with a spatially first order accuracy without the MUSCL. These results indicate that the difference in the stable regions between the Boris-HLLD and Boris-HLL solvers is attributed to the difference between the solvers.

\begin{figure}
\epsscale{1}
\plottwo{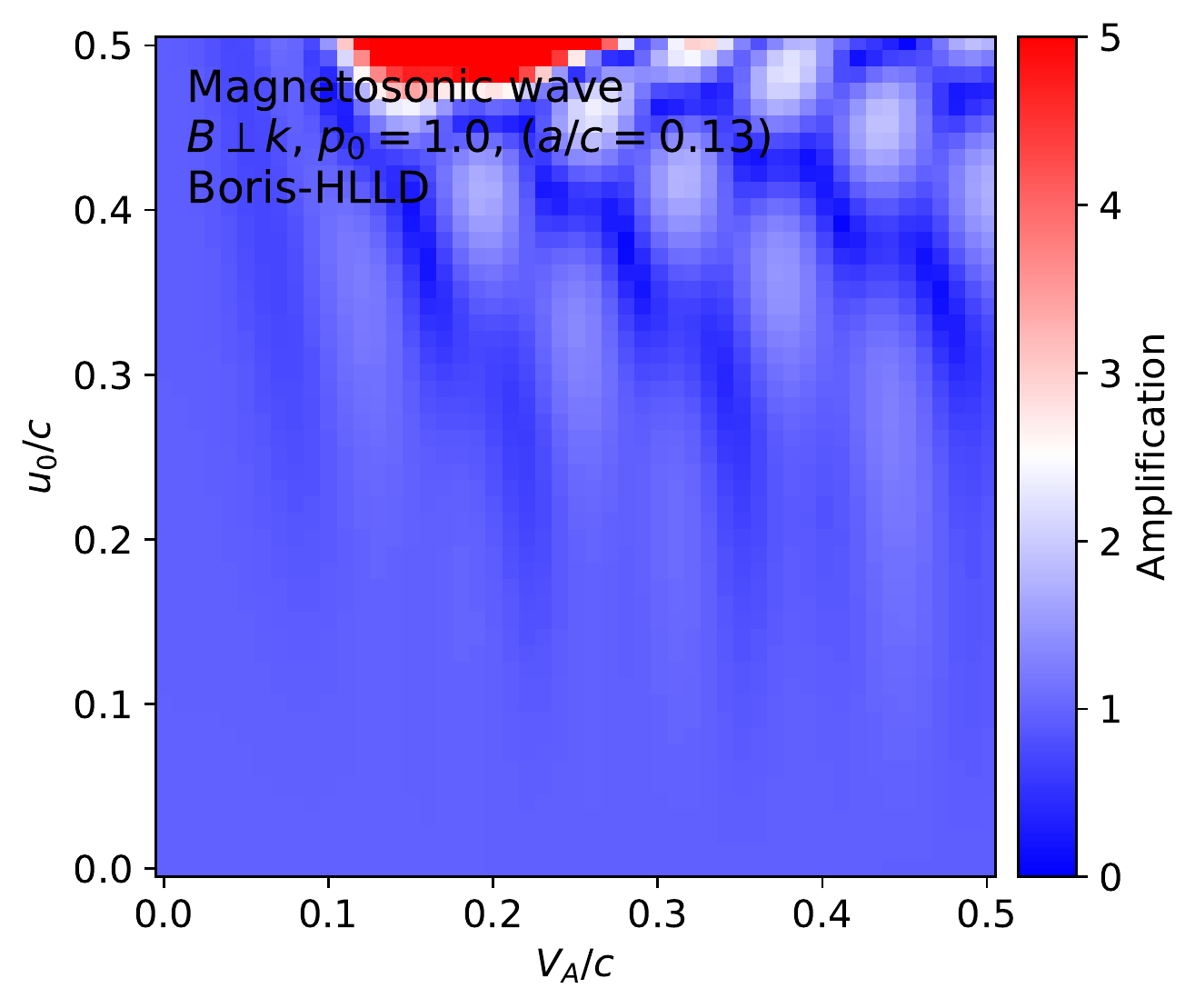}{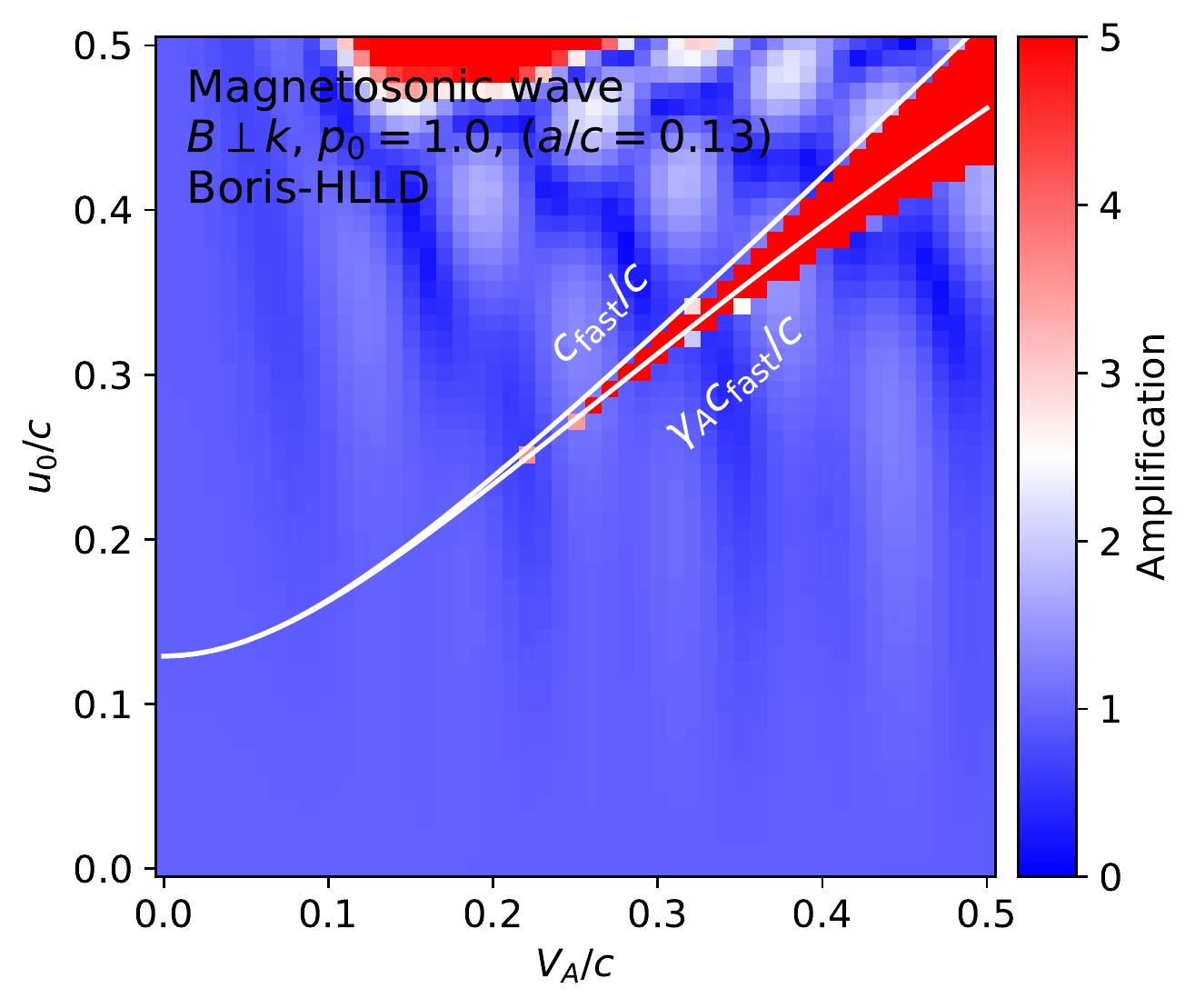}
\caption{
Magnification of the top left panel of Figure~\ref{map_alfvenwave.pdf}. The left and right panels, respectively, show the results obtained with and without the switching of the fast wave speed. The white lines in the right panel show the relationships of $u_0=c_\mathrm{fast}$ and $u_0=\gamma_A c_\mathrm{fast}$. 
\label{map_fastwave_switch.pdf}
}
\end{figure}

Figure~\ref{map_fastwave_switch.pdf} compares the stability of the magnetosonic wave between the schemes with and without the switching of the fast wave speed introduced in Section~\ref{sec:slsr}. Without the switching, instability arises for $c_\mathrm{fast} \lesssim u_0 \lesssim \gamma_A c_\mathrm{fast} $ (the red narrow region in the right panel). The regions above and below the lines represent super- and sub-magnetosonic flows, respectively. In the unstable region, the numerical flux $\mathbf{F}_L$ is adopted, which corresponds to a fully upwind difference. This numerical test indicates that this unstable region is stabilized by adopting the numerical flux for sub-magnetosonic flow $\mathbf{F}_L^*$. For the case of the Boris-HLL solver, we observe the same results as the case of the Boris-HLLD solver.

\subsection{Orszag-Tang vortex problem}
The Orszag-Tang vortex problem \citep{Orszag79} is widely used as a two-dimensional test problem for MHD schemes. The computational domain is $x,y \in [0, 2\pi]$ with $256^2$ mesh points. The periodic boundary conditions are imposed on the edges of the computational box, $x=0, 2\pi$, and $y=0, 2\pi$. The initial condition has distributions of  $\rho=\gamma^2$, $p=\gamma$, $u=-\sin y$, $v=\sin x$, $v=0$, $B_x=-B_0 \sin y$, $B_y = B_0 \sin 2 x$, $B_z=0$, and $B_0 = 1$.  In this problem, we set a pressure floor ($p_\mathrm{floor}=10^{-2}$) in order to detect the onset of numerical instability.

\begin{figure}
\epsscale{1.}
\plottwo{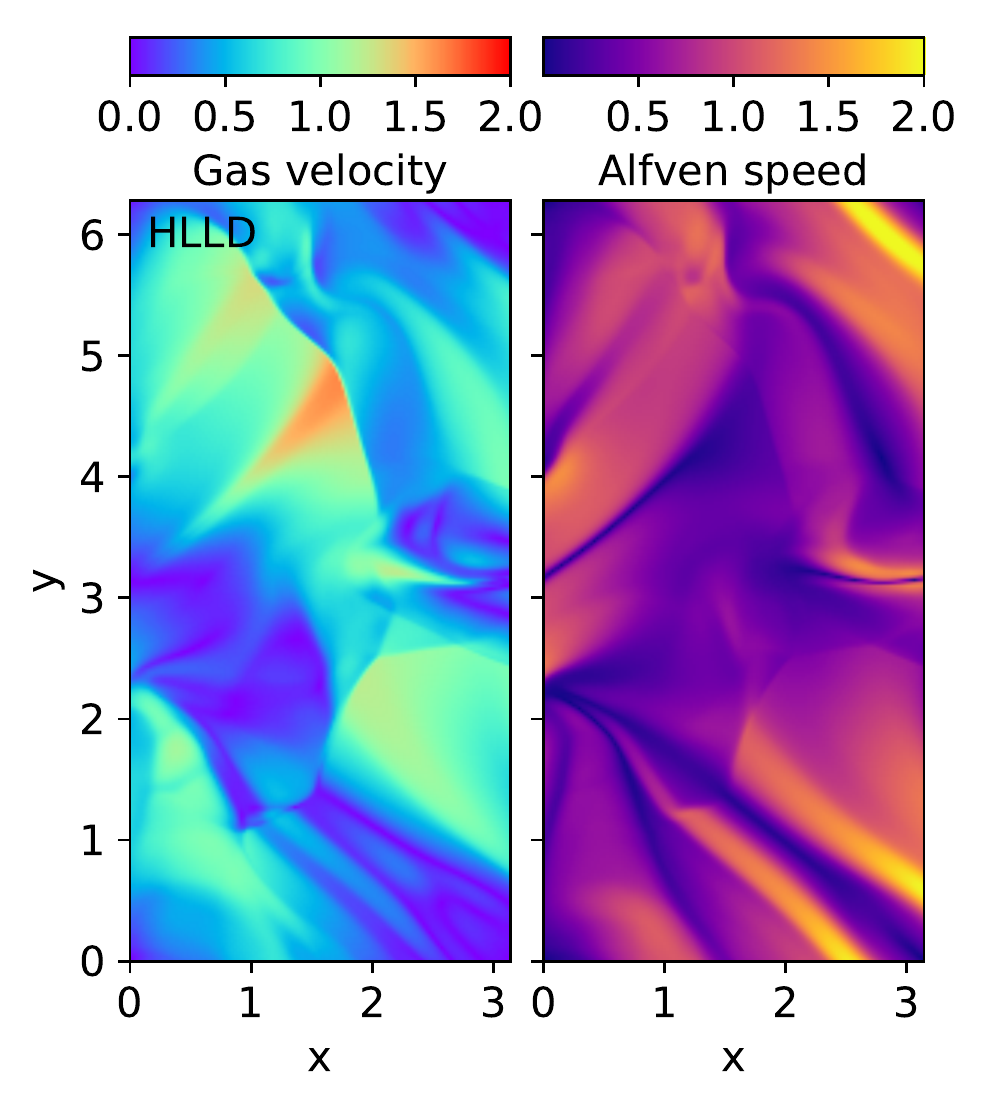}{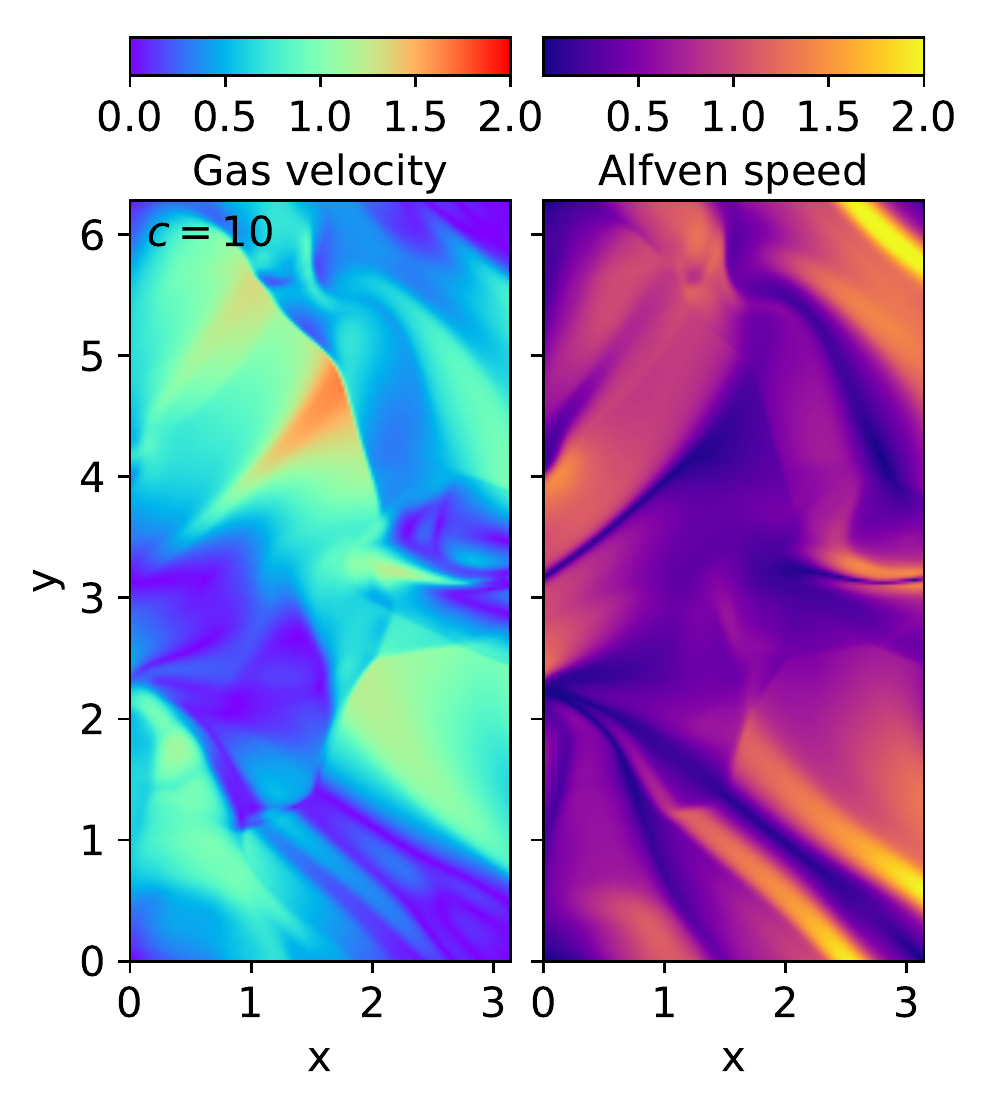}
\plottwo{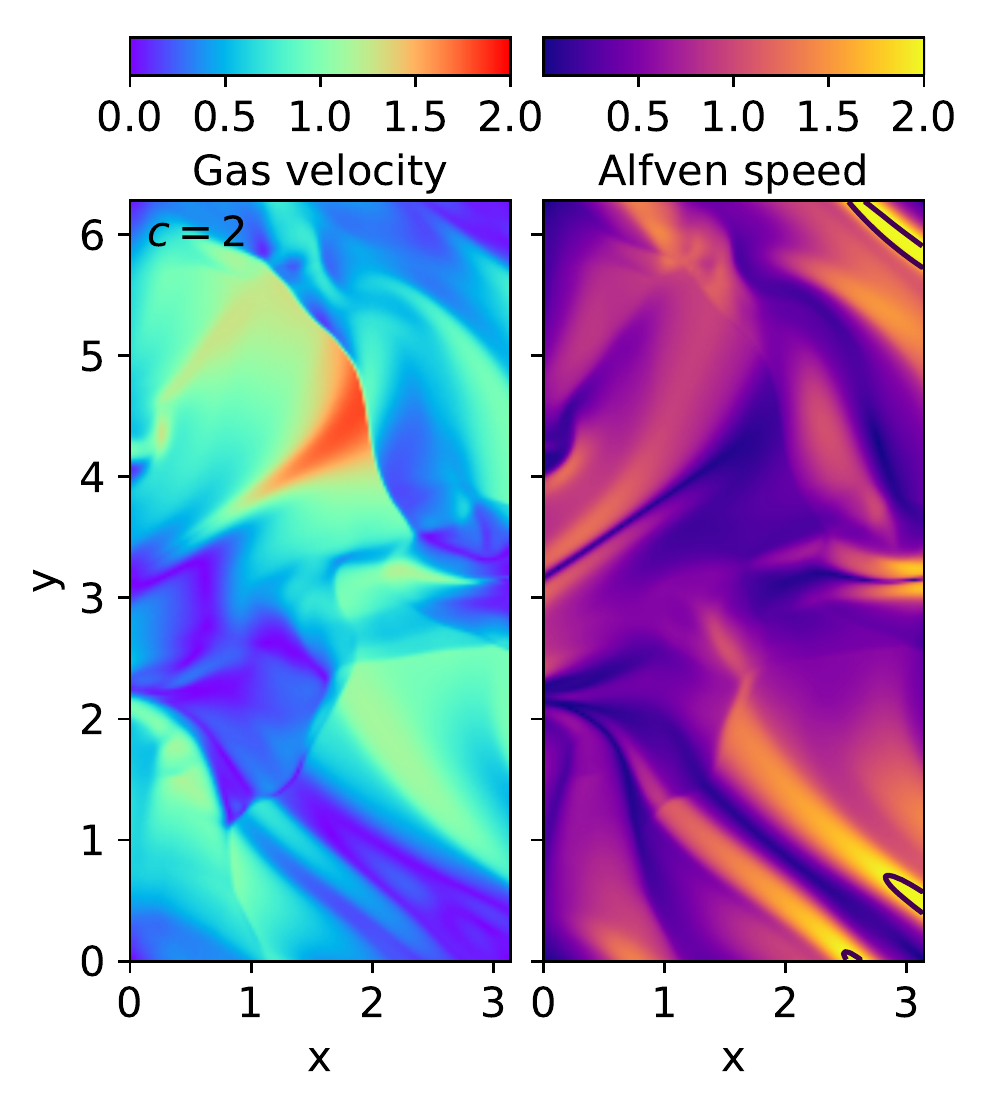}{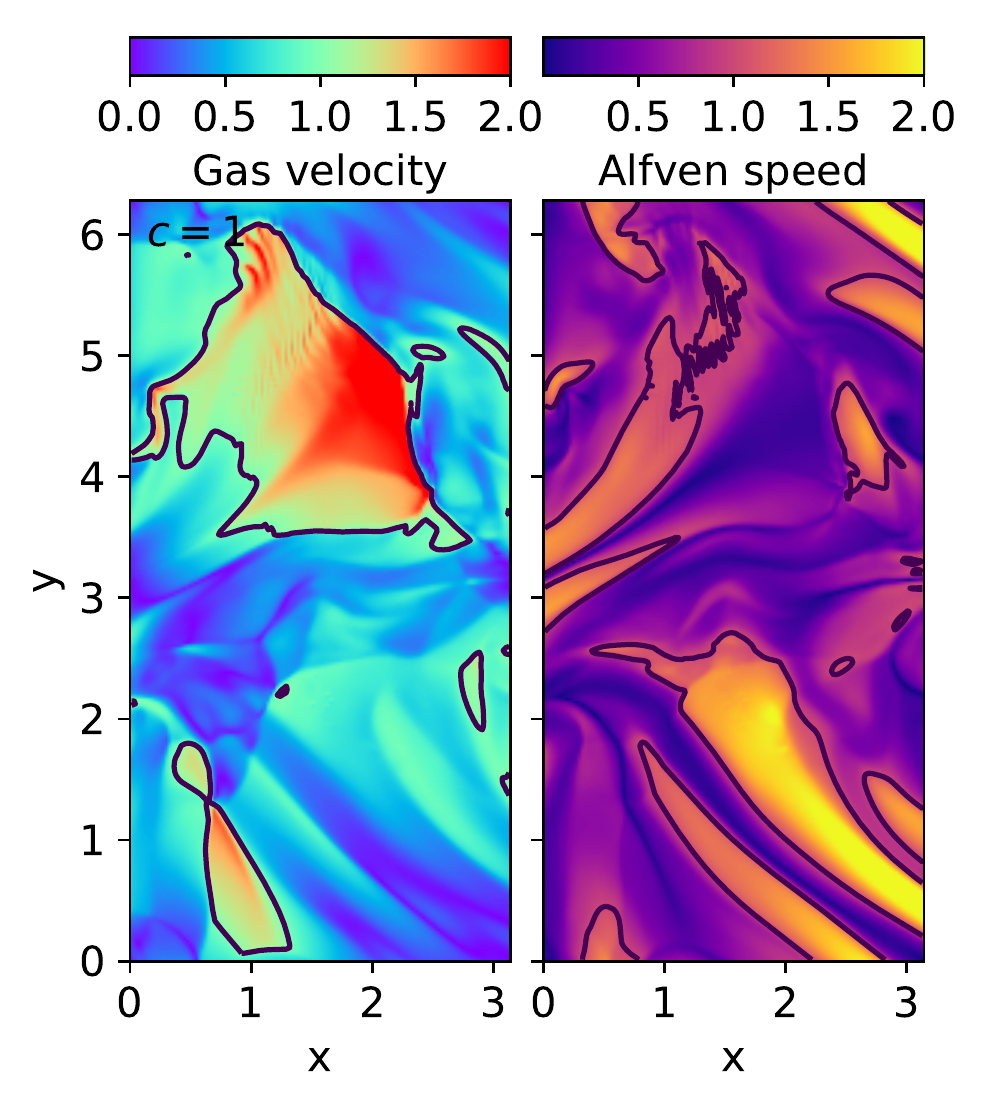}
\caption{
Orszag-Tang vortex problem solved by the HLLD and Boris-HLLD schemes with different speeds of light ($c=10, 2, 1$). The left and right panels show $|\mathbf{u}|$ and $V_A$, respectively, at $t=\pi$. The contours in the gas velocity map indicate the line where the gas velocity is equal to $c$, and the contours in the Alfv\'en speed map denote the line where the Alfv\'en speed is equal to $c$. \texttt{SFUMATO} was used for the calculations. 
\label{ozx.pdf}
}
\end{figure}

\begin{figure}
  \epsscale{0.5}
  \plotone{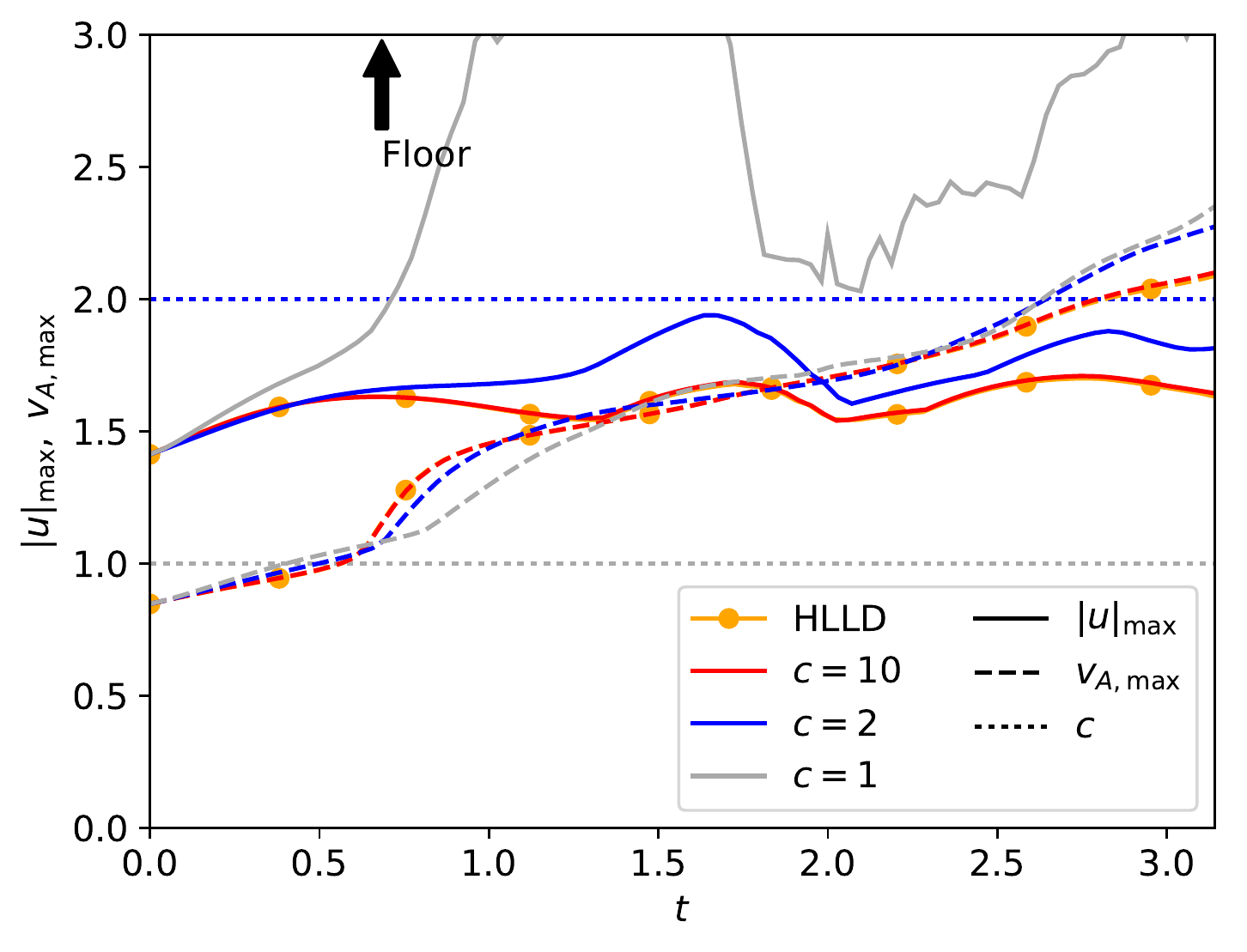}
  \caption{Maximum gas velocity (solid lines) and maximum Alfv\'en speed (dashed lines) as a function of time solved by HLLD and Boris-HLLD schemes with different speeds of light ($c=10, 2, 1$). The horizontal lines show the speeds of light $c$. The arrow indicates the time at which the pressure touches the floor for the first time for $c=1$.
  \label{oz_vb_t.pdf}}
\end{figure}

Figure~\ref{ozx.pdf} shows the gas velocity and Alfv\'en speed distributions obtained using the original HLLD scheme and the Boris-HLLD scheme with different speeds of light. The solution obtained with $c=10$ is quite similar to that of the HLLD scheme. The time evolutions for the two schemes coincide, as shown in Figure~\ref{oz_vb_t.pdf} (compare orange and red lines). For $c=2$ (blue lines), the Alfv\'en speed exceeds $c$ at the later stages. Even when the gas velocity is lower than $c$, the maximum velocity becomes larger than that in the HLLD solution. The distributions of gas velocity and Alfv\'en speed are slightly affected by $c$ (bottom left panel in Figure~\ref{ozx.pdf}).  For $c=1$, the gas velocity is higher than $c$ in a considerable area, in which numerical instability is prominent (bottom right panel in Figure~\ref{ozx.pdf}). Checkerboard-like instability appears in regions where the gas velocity exceeds $c$. The minimum pressure reaches the floor value in the early stage, as depicted by the arrow in Figure~\ref{oz_vb_t.pdf}. The maximum gas velocity grows exponentially at this time.

\begin{figure}
\epsscale{1.}
\plottwo{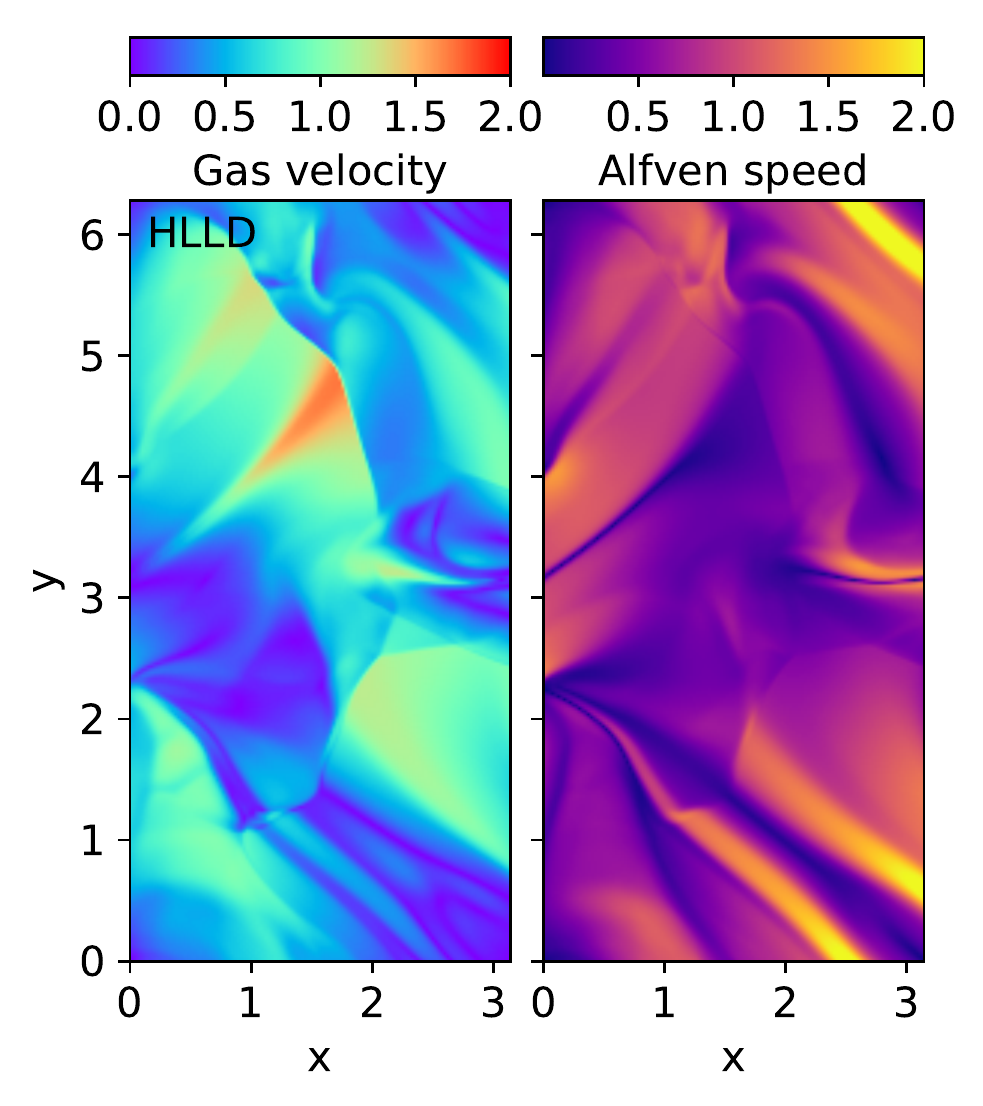}{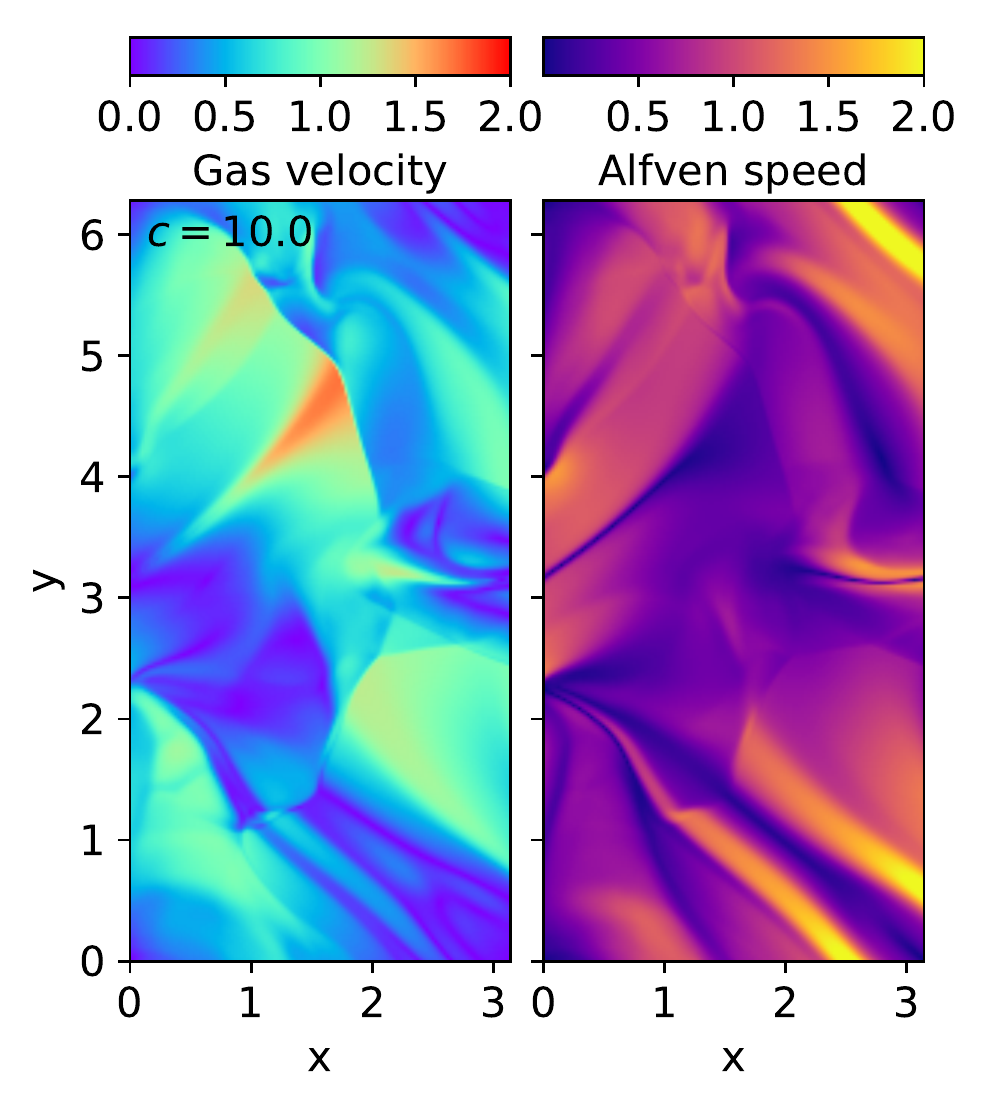}
\plottwo{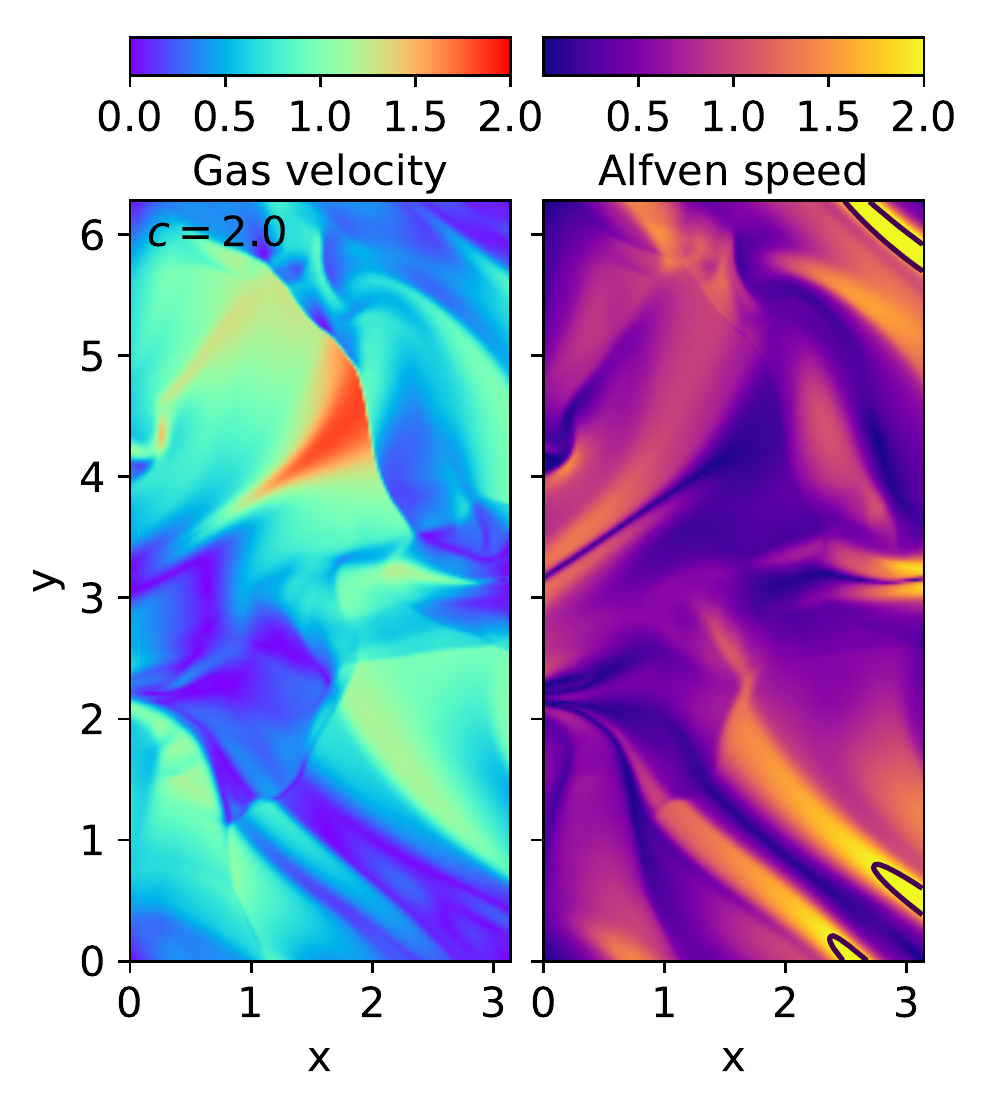}{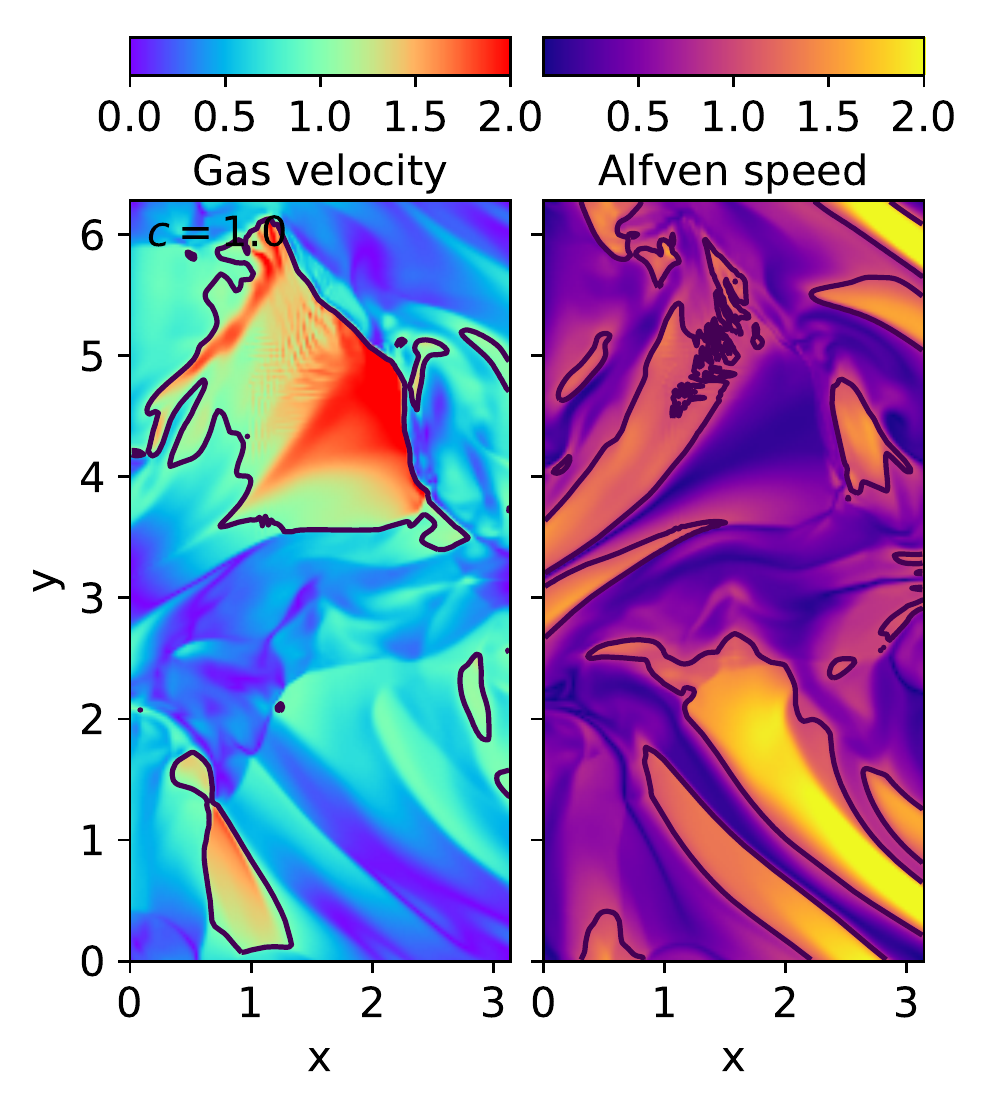}
\caption{
Orszag-Tang vortex problem solved by the HLLD and Boris-HLLD schemes with different speeds of light ($c=10, 2, 1$).  The left and right panels show $|\mathbf{u}|$ and $V_A$, respectively, at $t=\pi$. The contours in the gas velocity map indicate the line where the gas velocity is equal to $c$, and the contours in the Alfv\'en speed map denote the line where the Alfv\'en speed is equal to $c$. \texttt{Athena++} was used for the calculations.
\label{STozx.pdf}
}
\end{figure}

Figure~\ref{STozx.pdf} shows the same models as those in Figure~\ref{ozx.pdf} but calculated using \texttt{Athena++} for comparison. For $c=10$ and 2, the solutions obtained with the two codes are in agreement, indicating that the difference between the solutions with different $c$ values results from the Boris correction; it does not depend on the implementation of the code.  As mentioned, \texttt{SFUMATO} adopts the hyperbolic divergence cleaning method and \texttt{Athena++} adopts the constraint transport method for the treatment of $\nabla\cdot \mathbf{B}$. 
For $c=1$, checkerboard-like instability appears when the gas velocity exceeds $c$. Although numerical instability sometimes depends on  the implementation of numerical schemes, checkerboard-like instability appeared in the solutions obtained with \texttt{SFUAMTO} as well as those obtained with \texttt{Athena++}. This indicates that the numerical instability here arises from the basic equations adopted.

\section{Summary and discussion}
\label{sec:summary}

We proposed a high-resolution scheme for the ideal MHD equations that incorporates the simplified version of the Boris correction \citep{Gombosi02} into the HLLD Riemann solver. The Boris correction introduces an extra inertia term in the momentum equation.  The extra inertia is reflected by the magnetic field strength, and reduces the wave speeds. The wave speeds are bounded by the speed of light, which can be set to an artificially low value in order to avoid an extremely small timestep. As done by the original HLLD solver, the proposed scheme resolves four intermediate states separated by five waves: two fast waves, two Alfv\'en waves, and an entropy wave. In the limit of $V_A \ll c$, all the intermediate states and the numerical fluxes converge to those of the original HLLD.

Incorporating the Boris-HLLD scheme into existing code is simple. The numerical flux is replaced by that of the Boris-HLLD solver, 
the state vector is modified for the Boris correction,
and the CFL condition is modified so that the wave speed is multiplied by the factor $\gamma_A$.

We performed a stability analysis and showed the parameter space in which the scheme is stable. The scheme is stable when $|u| \lesssim 0.5c$ for a low Alfv\'en speed ($V_A \lesssim c$). For a high Alfv\'en speed ($V_A \gtrsim c$), the stable region becomes large for $|u| \lesssim (0.6-1) c$. The Boris-HLLD scheme shows larger stable regions than the Boris-HLL scheme. The scheme can be unstable even when $V_A < c$, and the semi-relativistic treatment is not necessary there. In this case, one can switch the scheme to the original HLLD scheme or adopt a sufficiently high value of $c$ to avoid instability. Practically, setting the speed of light to several times higher than the maximum gas speed is an acceptable compromise \citep{Rempel17}.

We showed the effects of the Boris correction on the solutions of non-steady-state problems (shock tube and the Orszag-Tang vortex problems). The Boris-HLLD scheme captures a contact discontinuity more sharply than the Boris-HLL scheme does. Although the semi-relativistic scheme including the Boris correction is powerful for stringent timestep problems, one has to check the impact of the modification, especially the dynamics in the region where $V_A \gtrsim c$. In other regions, the solution will be only weakly affected by the Boris correction. This scheme is therefore useful for avoiding an extremely high Alfv\'en speed in a relatively small volume in the computational domain. A conventional treatment for such a high Alfv\'en speed is to introduce a density floor \citep[e.g.,][]{Bai13}. However, with a density floor, mass and energy are unphysically injected into the computational domain. When self-gravity is taken into account, the influence of a density floor is more serious because it can increase gravity. The Boris-HLLD solver is an alternative method that overcomes these difficulties. 

\acknowledgments

Numerical computations were carried out in part on XC50 (ATERUI II) at the Center for Computational Astrophysics (CfCA), National Astronomical Observatory of Japan.
S.T. acknowledges support by the Research Fellowship of the Japan Society for the Promotion of Science (JSPS).
This research was supported by
JSPS KAKENHI Grant Numbers
15K04756,
16J02063,
17K05671,
17H02863,
17K05394,
18H04449,
18H05437, and
18K13579.

 \software{\texttt{SFUMATO} \citep{Matsumoto07}, \texttt{Athena++} \citep{Stone19}}

\end{document}